\documentclass[10pt,preprint]{aastex}
\usepackage{lscape}
\usepackage{subfigure} 

\accepted{}
\journalid{}{}
\articleid{}{}

\shortauthors{Stassun et al.}
\shorttitle{Improved physical properties of active low-mass objects}

\begin{document}

\def\ha{H$\alpha$}
\def\kms{km~s$^{-1}$}
\def\msun{M$_\odot$}
\def\rsun{R$_\odot$}
\def\lsun{L$_\odot$}
\def\mjup{M$_{\rm Jup}$}
\def\teff{$T_{\rm eff}$}
\def\logg{$\log g$}
\def\lbol{$L_{\rm bol}$}
\def\lha{$L_{{\rm H}\alpha}$}
\def\lx{$L_X$}

\def\2m{2M0535$-$05}

%\title{Improved Estimates of Physical Properties of Low-Mass Stars and Brown Dwarfs:
%An Empirical Correction for Suppressed Temperatures and Inflated Radii due to Activity}

\title{An Empirical Correction for Activity Effects on the Temperatures, Radii, and Estimated Masses 
of Low-Mass Stars and Brown Dwarfs}

\author{Keivan G.\ Stassun\altaffilmark{1,2},
Kaitlin M. Kratter\altaffilmark{3},
Aleks Scholz\altaffilmark{4}, and
Trent J.\ Dupuy\altaffilmark{3,5}}

\altaffiltext{1}{Department of Physics \& Astronomy, Vanderbilt University, Nashville, TN 37235, USA.  keivan.stassun@vanderbilt.edu}
\altaffiltext{2}{Department of Physics, Fisk University, Nashville, TN 37208, USA}
\altaffiltext{3}{Harvard-Smithsonian Center for Astrophysics, 60 Garden Street, Cambridge, MA 02138, USA}
\altaffiltext{4}{School of Cosmic Physics, Dublin Institute for Advanced Studies, 31 Fitzwilliam Place, Dublin 2, Ireland} 
\altaffiltext{5}{Hubble Fellow}

\begin{abstract} 
We present empirical relations for 
%\sout{correcting the estimated} \sout{quantitatively} 
determining the amount by which the 
effective temperatures and radii---and therefore the estimated masses---of low-mass stars and brown dwarfs
are altered due to chromospheric activity. 
%A significant fraction of M-dwarfs show evidence of chromospheric activity. Their measured physical properties are
%known to be discrepant with both their inactive counterparts, and theoretical models. 
%Here we derive an empirical relation that links both 
We base our relations
on a large set of low-mass stars in the field with \ha\ activity measurements, 
%and known distances, 
and on 
%We complement this sample with 
a set of low-mass eclipsing binaries with X-ray activity measurements
%and with directly measured temperatures and radii. 
%For these objects, we link the temperature suppression and radius inflation
%to the strength of X-ray emission and then 
from which we indirectly infer the
\ha\ activity. Both samples yield consistent relations
linking the amount by which an active object's temperature is suppressed, and its radius
inflated, to the strength of its \ha\ emission. 
%\ha\ emission to temperature suppression and radius inflation.
%\sout{Bolometric luminosity is found to be approximately preserved by these temperature and radius corrections.}
These relations are found to approximately preserve bolometric luminosity.
We apply these relations to the peculiar brown-dwarf
eclipsing binary \2m, in which the active, higher-mass brown dwarf has a
cooler temperature than its inactive, lower-mass companion. %\sout{We find that }
%\sout{the \ha-corrected
%temperatures bring the inferred masses of the brown dwarfs into agreement with theoretical isochrones}
The relations correctly reproduce the observed temperatures and radii of \2m\ after accounting for the \ha\ emission;
%\sout{our relations precisely explain the observed temperatures and radii of \2m\ from the observed \ha\ emission;}
\2m\ would be in precise agreement with theoretical isochrones were it inactive.
%\sout{Our empirical relations}
The relations that we present are applicable to brown dwarfs and low-mass stars with masses below 0.8 \msun\
and for which the activity, as measured by the fractional \ha\ luminosity, is in the range
$-4.6 \lesssim \log$~\lha/\lbol\ $\lesssim -3.3$. 
We expect these relations to be most useful for correcting
%\sout{temperatures and} 
radius and mass estimates of low-mass stars and brown dwarfs over their active lifetimes (few Gyr) 
%\sout{and thereby also the inferred masses of objects with unknown} 
and when the ages or distances (and therefore luminosities) are unknown. 
We also discuss the implications of this work for improved determinations of young cluster initial mass functions.
\end{abstract}

\keywords{stars: low-mass, brown dwarfs --- stars: fundamental parameters --- stars: activity}

\section{Introduction\label{intro}}
Observational evidence strongly indicates that the fundamental properties of low-mass stars 
can be altered in the presence of strong magnetic activity 
%that is common to low-mass objects
\citep[e.g.][]{morales08,lopez07,ribas06}.
In particular, observations of numerous active, low-mass eclipsing binary (EB)
stars have found the empirically measured stellar radii ($R$) to be inflated by $\approx$10\%, 
and the empirically measured stellar effective temperatures
(\teff) to be suppressed by $\approx$5\%, relative to the predictions of
standard theoretical stellar evolution models, which better match the properties of inactive objects
\citep[see][and references therein]{coughlin11,kraus11,morales10,torres10}.
Because the mass-radius and mass-\teff\ relationships are central to
our understanding of stellar evolution, resolving these discrepancies
will be critical to the ongoing development of accurate theoretical 
stellar models \citep[for a discussion, see][]{stassun10}. Accurate estimates of stellar radii are especially
important in the context of searches for transiting exoplanets,
which rely upon the assumed stellar radius/density to infer the planet radius/density.
 
%It is also important in a number of applied contexts
%to understand the nature and origin of these effects. 
%For example, accurate estimates of stellar radii are often 
%very important in the context of searches for transiting exoplanets, 
%which rely upon the assumed stellar radius/density to infer the planet radius/density.
%At the same time, estimates of stellar masses  are often derived from measurements of
%\teff\ together with an assumed $M$-\teff\ relation. 
%Consequently, the shape of the
%initial mass function (IMF) of stars in young clusters, particularly
%near and below the substellar boundary, may be significantly affected. 
%In particular, there will be a systematic tendency to underestimate the 
%masses of objects whose \teff\ have been suppressed. 

Activity effects also lead to errors in object masses ($M$) when these are derived from \teff.
%together with an assumed $M$-\teff\ relation.
A particularly salient example is \2m, an EB
in the Orion Nebula Cluster (age $\sim$1 Myr)
comprising two brown dwarfs \citep{stassun06,stassun07}.
The primary and secondary brown dwarf (BD) components of \2m\ have dynamically measured 
masses of 60$\pm$3 and 39$\pm$2 \mjup, respectively, 
and \teff\ ratio of $T_1 / T_2 = 0.952 \pm 0.004$ \citep{gomez09}.
That is, the 
system exhibits a reversal of the usual $M$-\teff\ relation, such that
the primary component is cooler than its companion.
This behavior is not predicted by theoretical models for coeval BDs. 
%{\bf (except in rare cases where deuterium fusion can heat up a lower mass BD)}.
Figure~\ref{fig:2m0535} shows the \2m\ system on the Hertzsprung-Russell (H-R) diagram 
compared to the 1~Myr isochrone of \citet{baraffe98}.
The secondary BD's \teff\ and bolometric luminosity (\lbol, calculated directly
from the empirically measured \teff\ and $R$) place it at a position
that is consistent with that predicted by the model isochrone.
In contrast, the primary is far displaced from its expected position, and so
{\it appears} to have a mass of only $\sim$25 \mjup---more than a factor of 2 
lower than its true mass---on the basis of its low \teff.
\citet{reiners07} used spectrally resolved H$\alpha$ measurements to show 
that, whereas the secondary BD in \2m\ is chromospherically quiet, the 
primary BD is highly chromospherically active, perhaps a consequence of its
rapid rotation \citep{gomez09}. Thus, magnetic activity in the primary BD
could be responsible for its highly suppressed \teff, similar to what has 
been seen for low-mass stellar EBs in the field. 

%Recent theoretical work suggests that the temperature suppresion might be caused by strong magnetic fields, which
%suppress convection, making heat transport less efficient \citep{chabrier07,macdonald2009}.
%In order to transport the same internal luminosity, the stellar radius $R$ inflates, leading to cooler \teff\ compared with the non-magnetic case.

Given the exhibition of activity-related \teff\ suppression in \2m, the 
first and only EB containing an active BD, it is 
likely that this phenomenon extends to other active BDs and low-mass stars in star 
forming regions and in the field.  This has important implications for 
estimating masses from the H-R diagram, as is often necessary at the youngest ages. When the distance of a low-mass main-sequence star is known, mass can be estimated from its luminosity either via empirical mass--magnitude relations or model mass-\lbol\ relations, circumventing the need to use \teff. However, the use of luminosities
can be problematic for very young objects, because they are  sensitive to age, accretion, disk
excess, and extinction---although careful evaluation of these parameters can mitigate these problems \citep[e.g.,][]{dario12}. For low-mass stars, \teff-derived parameters are useful on the pre--main-sequence, during which they evolve along nearly vertical tracks on the H-R diagram, and also while they remain active on the main-sequence lifetime. The activity timescale is relatively short at higher masses ($\lesssim$1~Gyr for spectral types $\leq$M2) but increases substantially at lower masses
($\gtrsim$7~Gyr for spectral types M5--M7) \citep{west08}.  
For substellar objects, determining physical properties is more complicated because such objects never reach a stable main sequence. Even with an accurately known distance and thus \lbol, the mass and age will be degenerate.  Thus, mass estimates for field BDs are generally inaccessible in the absence of age information.  
At the youngest ages ($\lesssim$100~Myr) the mass-\lbol\ relationship for BDs is substantially flatter as they are still undergoing significant contraction of their radii, so the situation is more akin to pre--main-sequence stars where it is important to know \teff\ accurately.
%As with low-mass stars, when \lbol\ and age are known they provide more reliable age estimates than \teff\ and age, given the current limitations in our understanding of substellar atmospheres. 
These evolutionary phases are also when BDs are most active, unlike at field ages when they have reached spectral types $\gtrsim$L4 for which activity is very rarely observed \citep[e.g.][]{burgasser02}.
%This is because 
%stars and BDs both proceed along nearly vertical tracks on the 
%H-R diagram as they contract to their final radii.  Thus, \teff\ must be 
%used to infer masses since \lbol\ is degenerate with age, which can be 
%very uncertain at such young ages. %\footnote{Note that at older ages, $\gtrsim$100~Myr, using \lbol\ and 
%age is generally preferred for estimating masses of BDs since this does not rely on
%\teff\ estimates from model atmospheres. {\bf [Remove this footnote.]}}  
%The case of the active primary in \2m\ serves as a warning that an unknown 
%fraction of substellar objects in young clusters and the field may have
%underestimated masses because 
%current evolutionary models do not take into account the ways in which magnetic activity
%may act to lower \teff\ and increase $R$. 
%%which appear to be substellar on the basis of their observed \teff\ or related
%%indicators (e.g., colors), may in fact be chromospherically active objects
%%with significantly higher masses.
%Indeed, {\bf at young ages (e.g., the Pleiades at 100 Myr), xx\% of low-mass stars are 
%\ha\ active}, and field stars with spectral types of M4 and later remain \ha\ active for 
%more than 4 Gyr \citep{west08}. 
%If \teff\ is lower than expected, then the resulting masses could be significantly underestimated.

Since magnetic activity seems to alter the fundamental properties of both stars and BDs,
it would be valuable to have an easily observable empirical metric with which to 
quantitatively assess the degree to which a given object's \teff\ has been suppressed and its
radius inflated. In the absence of a detailed understanding of the underlying physical 
causes of this effect, such a metric would be a useful ``stop-gap"---it mitigates one source
of uncertainty in estimates of fundamental parameters for very low mass objects. The desired 
metric will allow us to compare magnetically active objects with their inactive counterparts 
and with non-magnetic evolutionary tracks in order to improve the mass estimates for young low-mass stars and BDs.

%, and it would be ideal if this tool were relatively easy to observe
%for a typical low-mass star or BD. 
The aim of this paper is to derive such an empirical metric by relating the degree of \teff\ 
suppression and radius inflation to the strength of the \ha\ emission line, a commonly used and readily 
observable tracer of chromospheric activity \citep{scholz07,Berger:2006}.
%Because activity is thought to simultaneously cause an inflation in the stellar radii,
%our calibration also naturally relates \ha\ emission to the degree of $R$ inflation as well. 
A challenge with any activity-based measure is that most active objects exhibit
variability in their activity levels, and the amplitude of this variability can in some 
cases be quite large. 
For example, \citet{bell12} found variability of up to 30\% in \ha\ emission among a large sample of
M0--M9 dwarfs in the field on timescales of minutes to weeks.  
Thus, the calibration of \teff\ suppression and $R$ inflation to \ha\ emission requires
large statistical samples and/or calibration objects with highly accurate measurements in
order to identify robust mean relations. 
Fortunately for this paper, which is concerned with the most highly active young objects which may
experience the most significant \teff\ suppression and $R$ inflation, stronger \ha\ emitters also 
tend to be less time-variable, with the typical \ha\
variability being less than 10\% for the strongest \ha\ emitters \citep{bell12}.

In Sec.~\ref{methods} we describe our approach and the data 
we use to establish the empirical relationships of
$\Delta$\teff\ and $\Delta R$ vs.\ \lha/\lbol. In Sec.~\ref{results}
we present the resulting relations and apply them to \2m\ as a test case,
finding that the position of the chromospherically active primary BD in the H-R diagram 
%is effectively ``corrected" to its expected position
can be fully explained as an offset from its theoretically expected position, due
to the effects of activity on its $R$ and \teff\ 
(Fig.~\ref{fig:2m0535}). In Sec.~\ref{summary} we discuss
the broader application of these relations to other low-mass objects, and 
their possible ramifications for the inferred IMFs of young clusters. 
We conclude with a summary in Sec.~\ref{conclude}.

\section{Methods and Data Used \label{methods}}
Our aim is to empirically determine a relationship between the temperature 
suppression (as compared to evolutionary models) and the level of activity 
as measured from \ha.  We adopt a simple linear form for this relationship:
%{\bf [Note new formulae with offset of +4]}
\begin{equation}
\label{eq:Trelation}
\Delta T_{\rm eff} / T_{\rm eff} = m_T \times (\log L_{{\rm H}\alpha} / L_{\rm bol} +4) + b_T
\end{equation}
where $\Delta T_{\rm eff} / T_{\rm eff}$ is the fractional temperature offset 
(observed minus model), \lha/\lbol\ is the ratio of luminosity in the \ha\ line
to the bolometric luminosity, $m_T$ and $b_T$ are linear coefficients,
and the offset of $+$4 is for convenience given the typical 
value of $\log$~\lha/\lbol\ $\approx -4$ for our study sample (see \S\ref{results}).
We determine a similar relationship for the fractional radius inflation:
\begin{equation}
\label{eq:Rrelation}
\Delta R/R = m_R \times (\log L_{{\rm H}\alpha} / L_{\rm bol} +4) + b_R .
\end{equation}
In the following analysis, we use two different samples and approaches
to determine the linear coefficients of Eqs.~\ref{eq:Trelation}
and \ref{eq:Rrelation}.  In the first approach we use a large sample 
of stars without direct mass, radius, or \teff\ measurements but 
with direct distances, spectral types, and reliable \ha\ measurements.  
The second approach uses the much smaller sample of stars in EBs that have directly measured 
masses, radii, and reliable \teff, but for which we must use X-ray flux as a proxy for \ha.

\subsection{Nearby M Dwarfs with \ha\ Emission\label{mdwarfs}}
We first consider the large set of nearby field M dwarfs with well measured spectral 
types and \ha\ equivalent widths (EWs) from the Palomar/Michigan State University
catalog \citep[PMSU;][]{reid95,hawley96}. 
In particular, following \citet{morales08}, we restrict 
ourselves to the sample of 746 stars with distances determined directly from 
trigonometric parallaxes. These distances allow \lbol\ to be calculated from the 
observed 2MASS $K_S$ magnitudes together with $K$-band bolometric corrections 
from \citet{bessell98}. The catalog reported spectral types are transformed to 
\teff\ according to \citet{leggett96} and \citet{bessell91}. 
%\textbf{Keivan: \citet{leggett96} appears to only include M1--M9 or so, so this couldn't be the entire SpT-to-\teff\ used, right?}
From these \teff\ and \lbol\ \citep[][their Tables 1 and 2]{morales08}
we also calculate the stellar radii, $R$, according to the usual Stefan-Boltzmann relation. 
In order to isolate a consistent sample of main-sequence stars 
as representative as possible of Galactic disk population stars,
we removed all stars identified by \citet{morales08} as being unresolved binaries, members of young moving groups,
part of the old halo population, or evolved subdwarfs. 
The sample of 746 nearby stars with trigonometric distances is already dominated by field-age disk stars, so 
these cuts cause only 54 stars to be removed from the sample. 
We further restricted the sample to those with $M<1$ \msun, leaving 669 stars in our sample. 
Finally, we use the solar-metallicity isochrone of \citet{baraffe98} with mixing-length $\alpha$=1.0
to interpolate the predicted stellar masses ($M$), \teff, and $R$ as functions of \lbol.
Our results below do not change significantly for adopted isochrone ages in the range 1--5 Gyr, typical of field star 
ages for the Galactic disk population, so we adopt the isochrone at 3~Gyr throughout for simplicity.

\begin{figure}
%\epsscale{0.5}
\includegraphics[width=2.5in,angle=90]{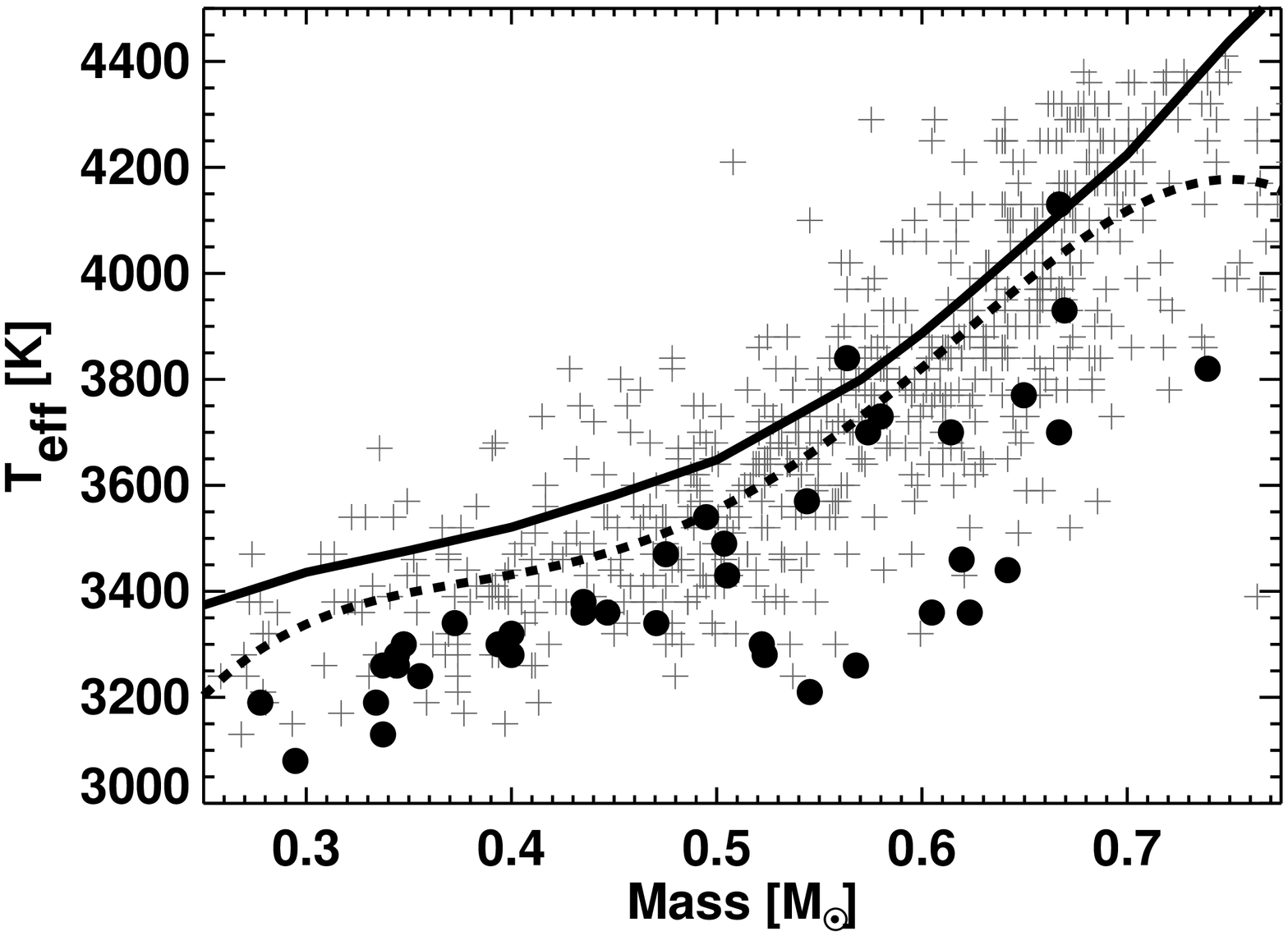}
\includegraphics[width=2.5in,angle=90]{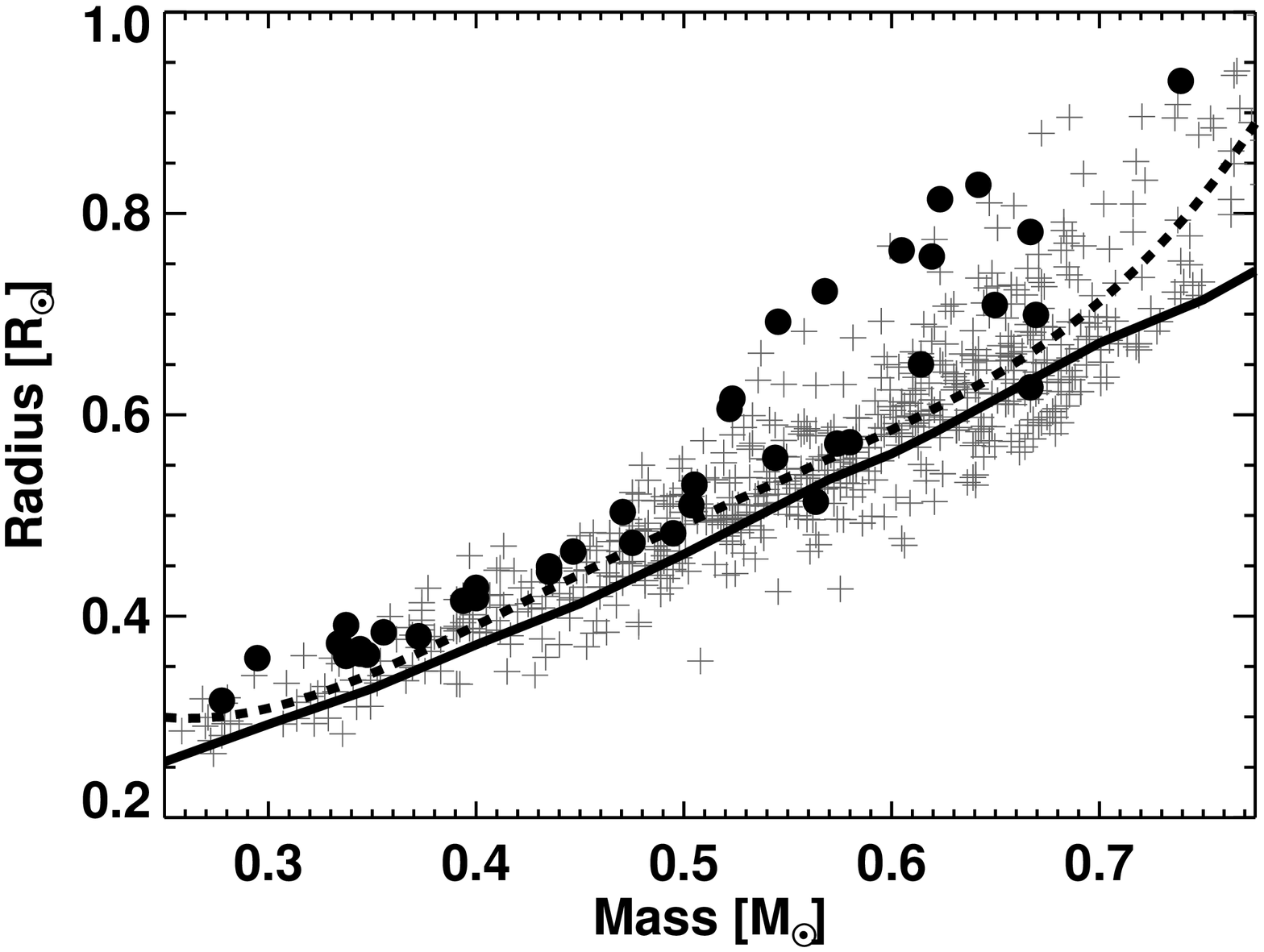}
\caption{\label{fig:pmsudata} 
Effective Temperature vs.\ Mass (left) and Radius vs.\ Mass (right) for M-dwarfs with 
trigonometric distances and \ha\ measurements from the PMSU catalog. 
Active objects (i.e., those with \ha\ in emission) are represented by filled symbols.
For reference, the 3 Gyr theoretical isochrone of \citet{baraffe98} is represented in
both figures as a solid curve. A polynomial fit to the non-active objects is represented 
by a dashed curve. The \ha-active dwarfs are found to be significantly displaced to lower
\teff\ and larger radii as compared to both the theoretical isochrone and the non-active dwarfs.
}
\end{figure}

Fig.~\ref{fig:pmsudata} shows the estimated \teff\ and $R$ of the PMSU sample stars
as a function of $M$.
Strongly chromospherically active stars---defined here as those with \ha\ in emission---show
a clear displacement to lower \teff\ relative to both the theoretical isochrone
and to the non-active stars, whereas non-active stars more closely track the isochrone.
The \ha\ active stars also show a displacement to larger $R$ relative to both the 
theoretical isochrone and to the non-active stars. 
The mean offset of the \ha\ active stars relative to the isochrone (solid curve) 
is 10.0$\sigma$ for $R$ and $-11.1\sigma$ for \teff, where
$\sigma$ is the standard deviation of the mean (i.e., r.m.s./$\sqrt{N}$). 
The mean offset relative to the non-active stars (dashed curve) 
is 6.8$\sigma$ for $R$ and $-7.5\sigma$ for \teff. 
%{\bf [Trent's note: I didn't really get the value of quoting
%differences in sigma between active and nonactive stars.  It might be
%misleading for some readers, since it could imply that we have used
%measurement errors to compute these, but it's just rms/sqrt(N) (and we
%actually never even assign errors to the mass, radius, and Teff estimates).]} 
The non-active stars themselves show a mild displacement to lower \teff\ and larger $R$
relative to the theoretical isochrone. 
We note that at least some of the ``non-active" stars in
the sample may in fact possess mild chromospheric activity. Mildly active M-type
stars can exhibit \ha\ in {\it absorption} \citep[e.g.][]{walkowicz09,west11}
and thus would not be identified as strongly
active according to our \ha\ emission criterion but could still manifest mild
activity-related effects. 
Alternatively, these offsets for the non-active stars may simply indicate 
%a real systematic error in the theoretical isochrone. 
mild systematics in the transformation from the observed $K$-band magnitudes into
masses and radii via bolometric corrections; 
indeed, the polynomial fits to the non-active stars closely parallel the theoretical isochrone
(except at the highest masses; see also Sec.~\ref{sec:relations}). 
%suggesting that this is likely to be the primary cause. 
In what follows we conservatively
measure the \teff\ and $R$ offsets of the \ha\ active stars relative to the non-active
stars, effectively using the latter to 
%define empirical $M$--\teff\ and $M$--$R$ relations.
calibrate the derived stellar masses and radii to the theoretical isochrone. 
For reference, the polynomial fits with which we describe the non-active stars 
(dashed curves in Fig.~\ref{fig:pmsudata}) are:
$R$/\rsun\ = $\sum_{i=0}^{4} r_i (M$/\msun)$^i$ and 
\teff/K = $\sum_{i=0}^{4} t_i (M$/\msun)$^i$, with
$r_i = \{1.3836, -10.9203, 37.7554, -52.3712, 26.8055\}$ and
$t_i = \{-1384.1, 41857.5, -136147, 192515, -96288.4\}$, applicable for $M < 0.8$ \msun.

\begin{figure}
\subfigure[]{
\includegraphics[width=2.5in,angle=90]{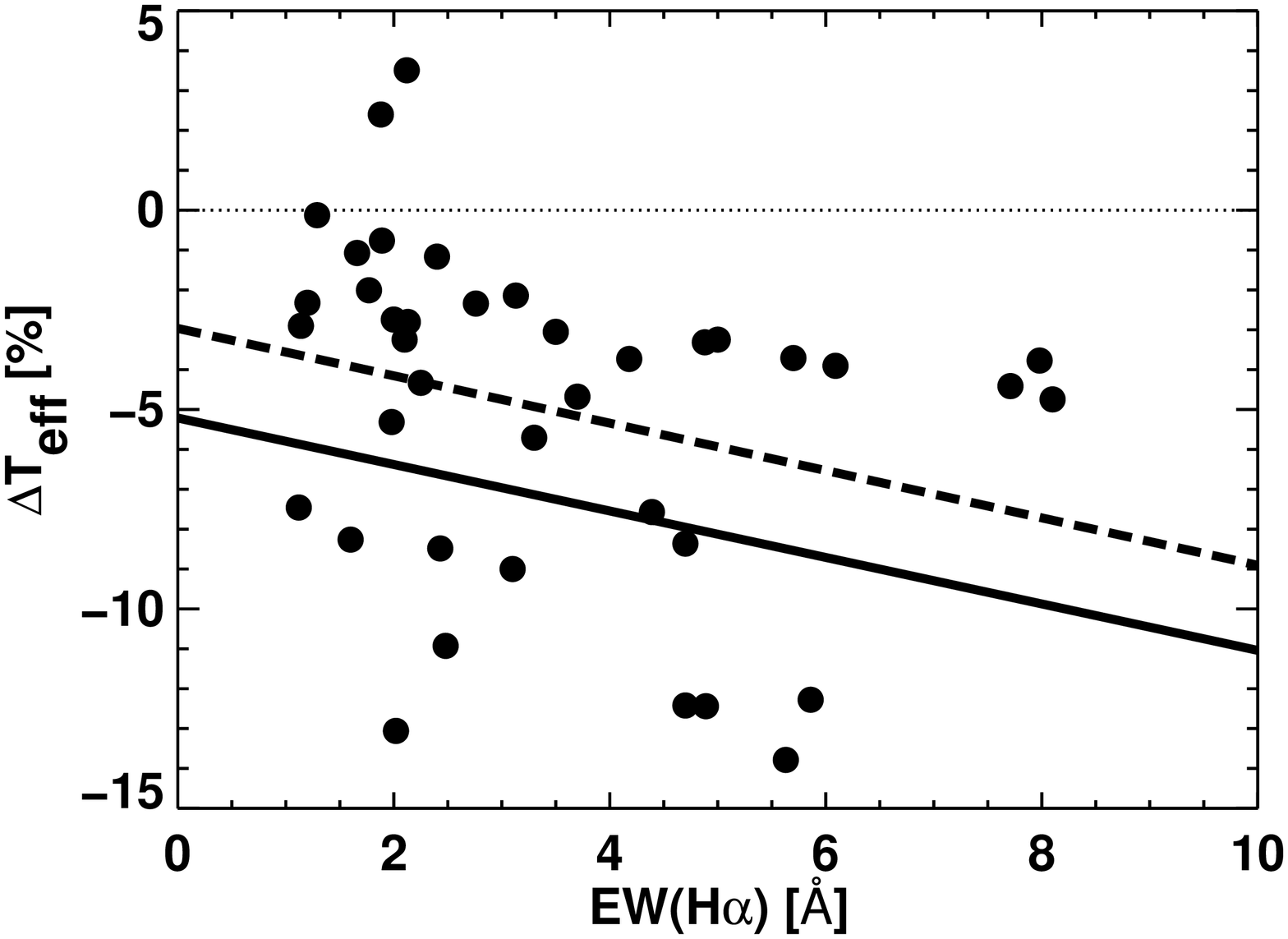} 
\includegraphics[width=2.5in,angle=90]{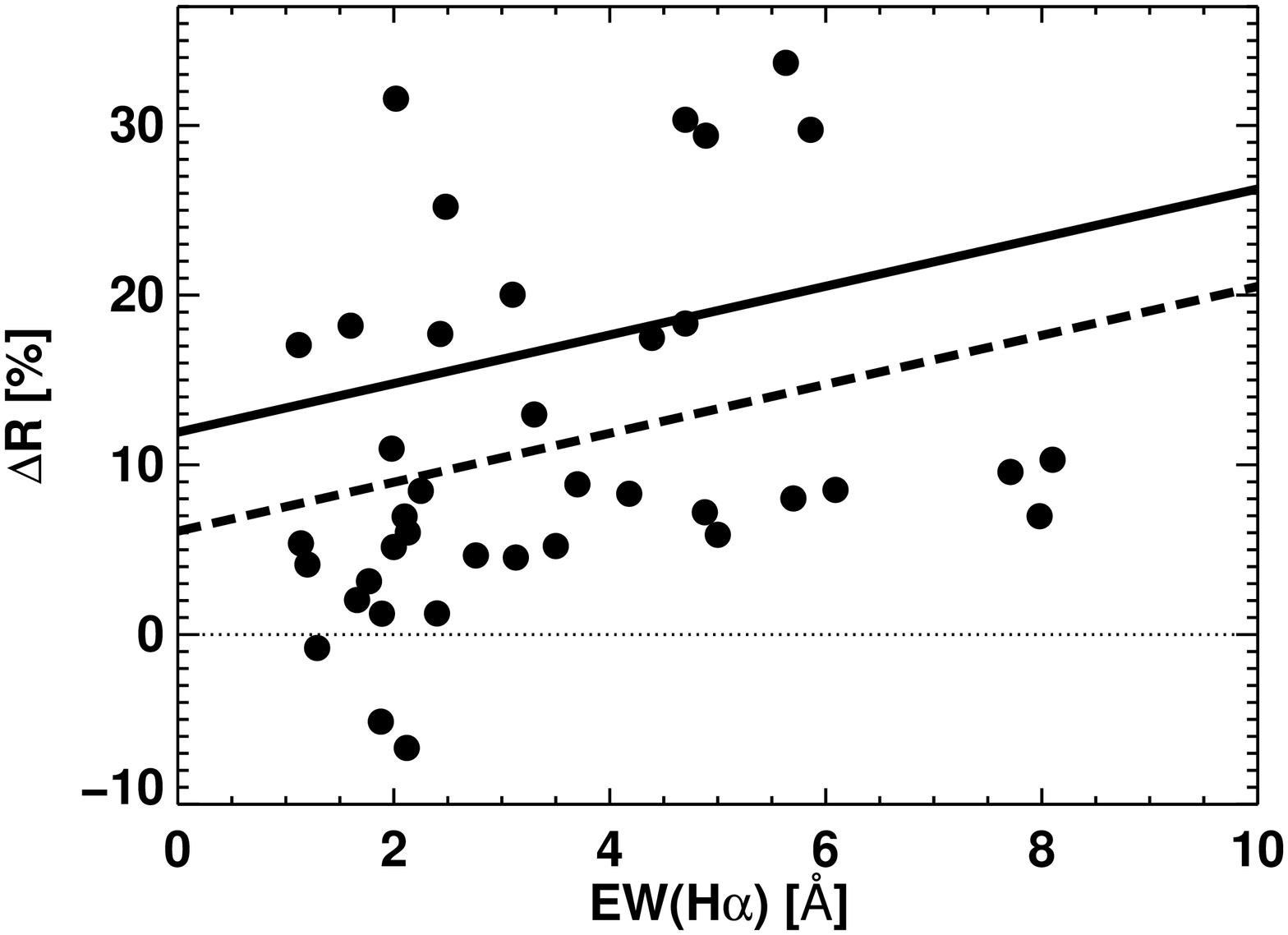}
\label{fig:pmsuewhaeffect}
}
%\caption{
%}
%\end{figure}

%\begin{figure}
\subfigure[]{
\includegraphics[width=2.5in,angle=90]{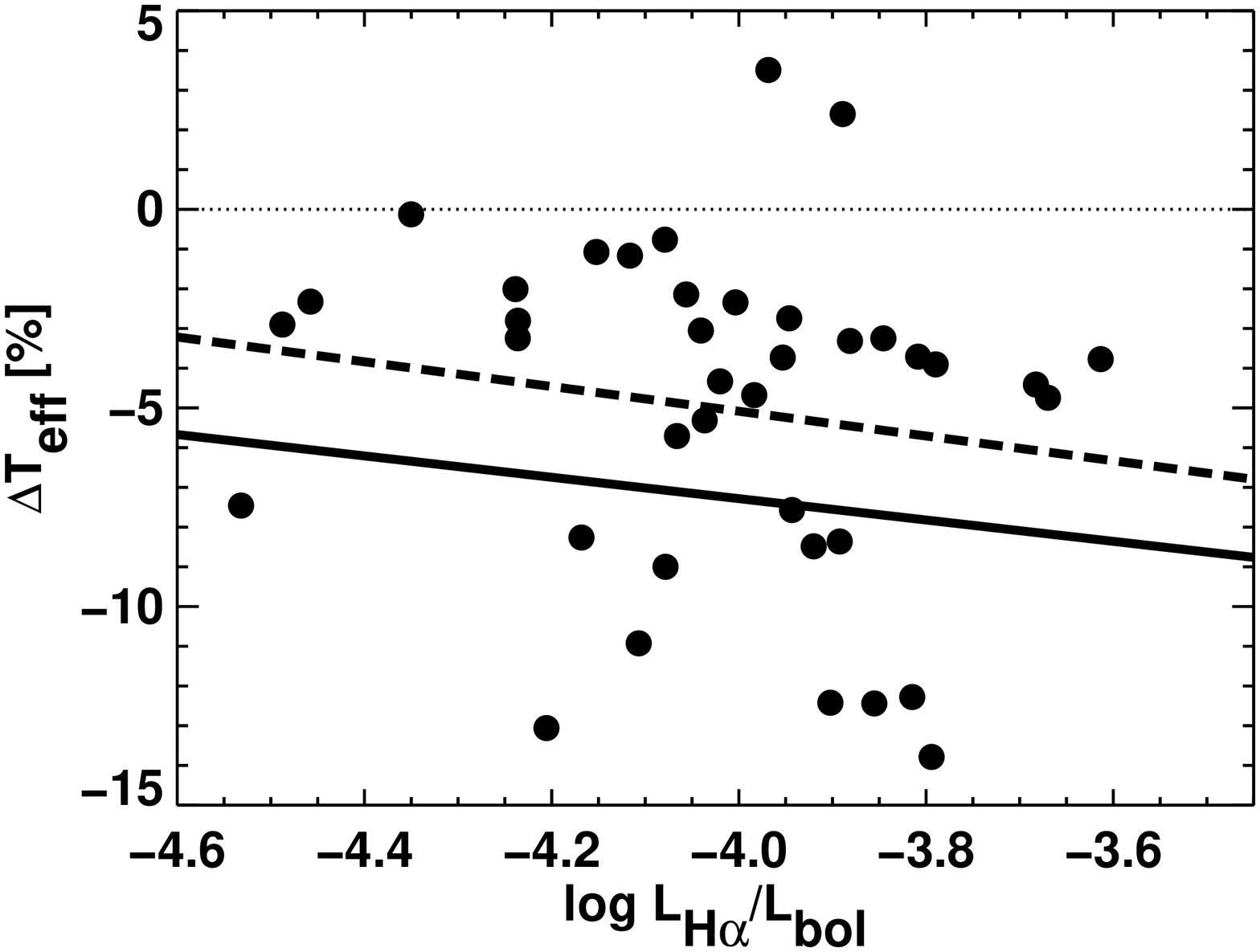} 
\includegraphics[width=2.5in,angle=90]{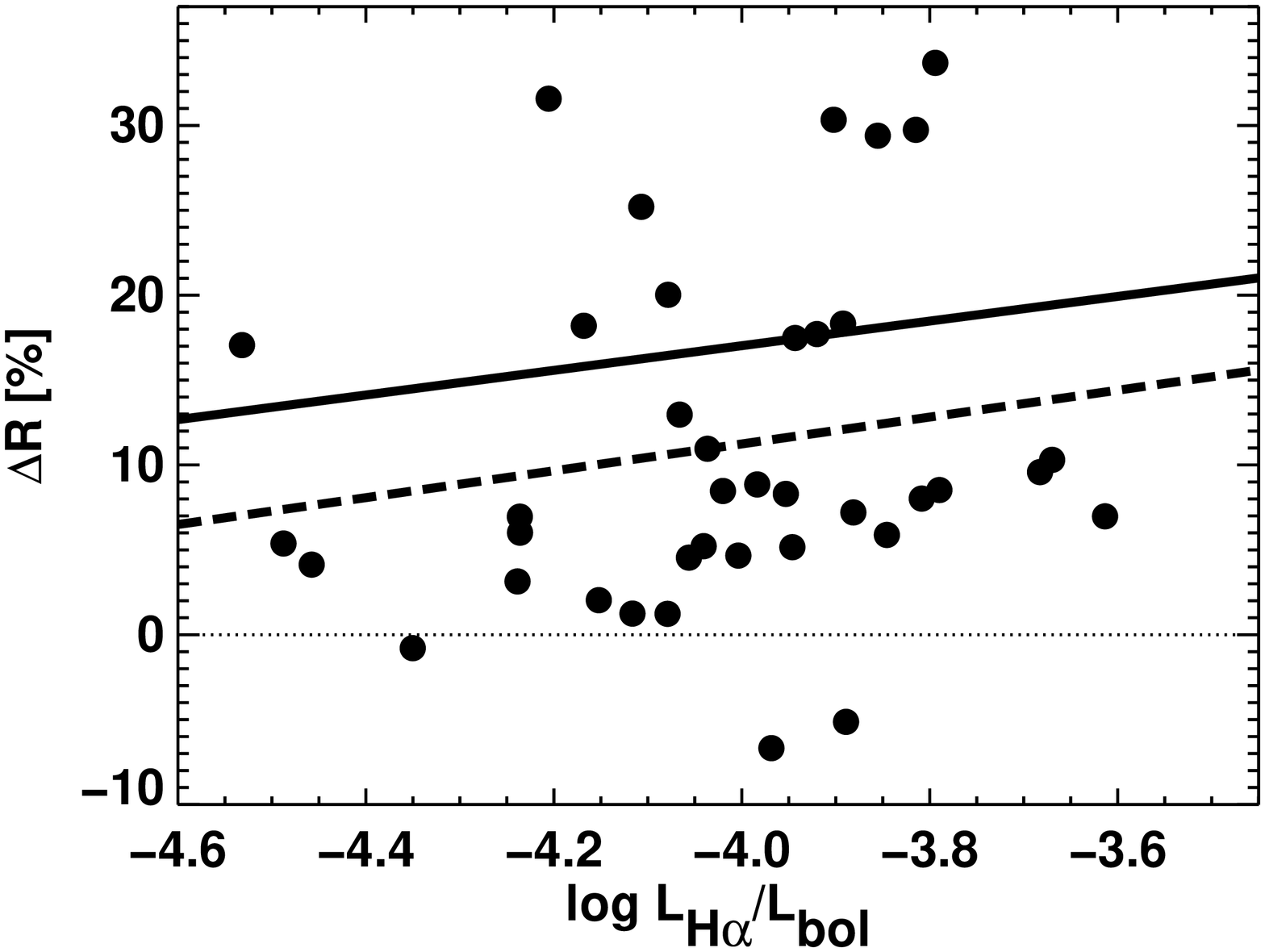}
\label{fig:pmsurelations}
}
\caption{Suppression of effective temperatures (left) and inflation of radii (right) as a function of
\ha\ emission strength for the \ha-active sample from Fig.~\ref{fig:pmsudata} (filled symbols),
measured relative to the non-active stars. 
In (top), the abscissa is the directly measured \ha\ equivalent width, in (bottom) the \ha\ equivalent
widths have been converted to fractional \ha\ luminosity. 
In all panels, the solid line is a linear fit to the \teff\ and radius differences relative to the
theoretical isochrone (solid curves in Fig.~\ref{fig:pmsudata}); 
the dashed line is a linear fit to the \teff\ and radius differences relative
to the non-active stars (dashed curves in Fig.~\ref{fig:pmsudata}). 
For our analysis we use the more conservative \teff\ and radius differences measured relative
to the non-active stars (shown here as filled symbols and dashed lines).
}
\end{figure}

Fig.~\ref{fig:pmsuewhaeffect} shows the offsets in estimated \teff\ and $R$ for the \ha\ active sample stars
%relative to the non-active stars 
as a function of the observed EW(\ha). We fit a simple 
least-squares linear relation to each (dashed lines), yielding the following:
$$ \Delta R/R\, [\%] = (1.4 \pm 0.8) \times {\rm EW(H}\alpha)[{\rm \AA}] + (6.1 \pm 3.2) $$
$$ \Delta T_{\rm eff}/T_{\rm eff}\, [\%] = (-0.6 \pm 0.3) \times {\rm EW(H}\alpha)[{\rm \AA}] + (-3.0 \pm 1.3) .$$
For example, for EW(\ha)=4\AA, the relation predicts a \teff\ suppression of $\approx$5\% and a radius
inflation of $\approx$11\%. There is substantial scatter in the data, nonetheless a Kendall's $\tau$ 
correlation test gives a null-hypothesis probability of only 0.9\% and 0.7\% that the quantities are not 
correlated in $\Delta R$ and in $\Delta$\teff, respectively, 
indicating that the correlations are statistically significant at $>$99\% confidence. 
For comparison, a linear fit to the \teff\ and $R$ offsets calculated relative to the 
theoretical isochrone is shown as a solid line. The linear fit cofficients are nearly identical
to those above, except of course for a larger mean offset in both quantities. 

To calculate the \ha\ luminosity, \lha, from the EW(\ha) measurements for Equations~\ref{eq:Trelation}--\ref{eq:Rrelation},
we determined scaling factors from a set of stellar atmosphere 
models for \teff\ ranging from 3000 to 5000\,K in steps of 100\,K 
and $\log g = 4.0$. We use the STARdusty1999 model spectra, which are based on the 
NextGen models updated with new H$_2$O and TiO opacities \citep{allard00}. We 
obtain the continuum flux around the \ha\ line (6500--6600~\AA)
%in flux units per surface element 
%and the luminosity for this surface element 
for each \teff, 
%The quotient $q$ of the continuum flux relative to the luminosity \lbol\ is the required scaling factor; 
%To be able to carry out this conversion for objects with a 
%given \teff\ 
and we fit the flux--\teff\ relation with a third-order polynomial.
The measured EW(\ha) is then multiplied by the \ha\ continuum flux from this relation, and multiplied by the 
surface area of the star using the measured $R$ calculated as described above, thus giving the total \lha.
%the EW(\ha) value is multiplied by this factor to obtain \lha/\lbol. 
%{\bf Note from Aleks: the models do not include the magnetic effects, i.e. their
%effective temperatures are off for magnetically active objects. Since Lbol is derived
%from Teff here, this leads to a problem. I don't see an easy way out, for this conversion
%to work we either need to assume Teff or the radius. So, the conversion is not entirely
%consistent, maybe we should explicitly state that, while still using this LHa/Lbol
%as activity indicator.}
Fig.~\ref{fig:pmsurelations} shows the same data as in Fig.~\ref{fig:pmsuewhaeffect}, but now with
EW(\ha) converted to \lha/\lbol. 
The resulting relations between the \teff\ and $R$ offsets vs.\ \lha/\lbol\ are 
discussed in Sec.~\ref{results}.
%{\bf [Aleks, I modified this paragraph to reflect what we actually did. Note that we did *not* get Lbol
%from the models! So I think we are ok. OK? -- Oh, great, I wasn't aware that you used the actual radii, 
%that's much better. We still get the continuum flux around Halpha from the non-magnetic models, but I don't see
%how this could introduce a systematic error. ]} 

\subsection{Low-mass Eclipsing Binary Stars with X-ray Emission}
In the second approach, we use the small set of low-mass EBs with accurately
measured $M$, $R$, \teff, and X-ray luminosities (\lx) from 
\citet{lopez07}.\footnote{We use the ``case~1" \lx\ values from \citet{lopez07}. The results
do not change significantly if we adopt the ``case~2" or ``case~3" \lx\ values instead.}
The sample includes 11 individual stars in 7 EB systems spanning the range 
\teff=3125--5300~K and $M$=0.21--0.96~\msun. 

\begin{figure}
%\epsscale{0.5}
\subfigure[]{
\includegraphics[width=2.5in,angle=90]{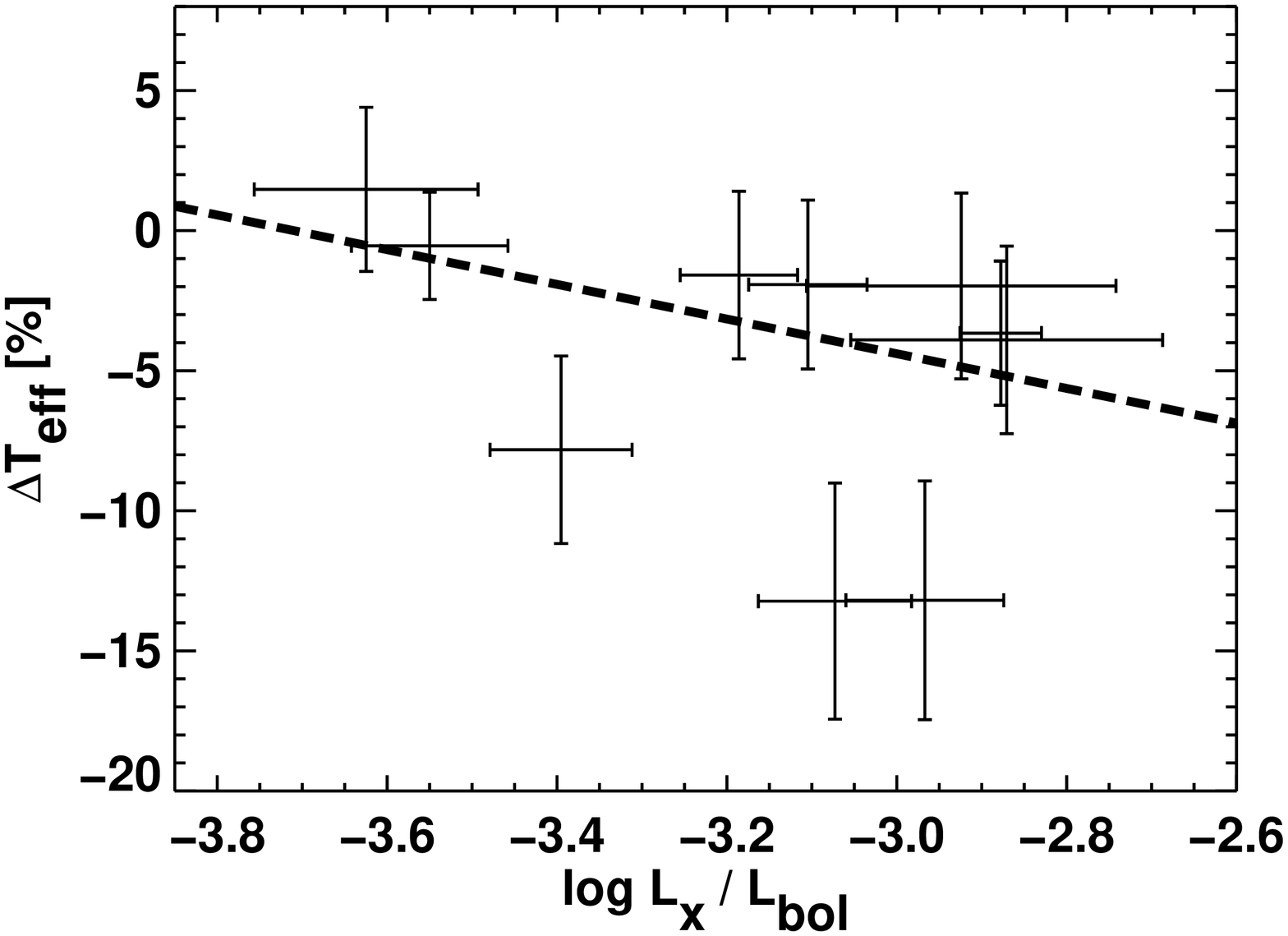}
\includegraphics[width=2.5in,angle=90]{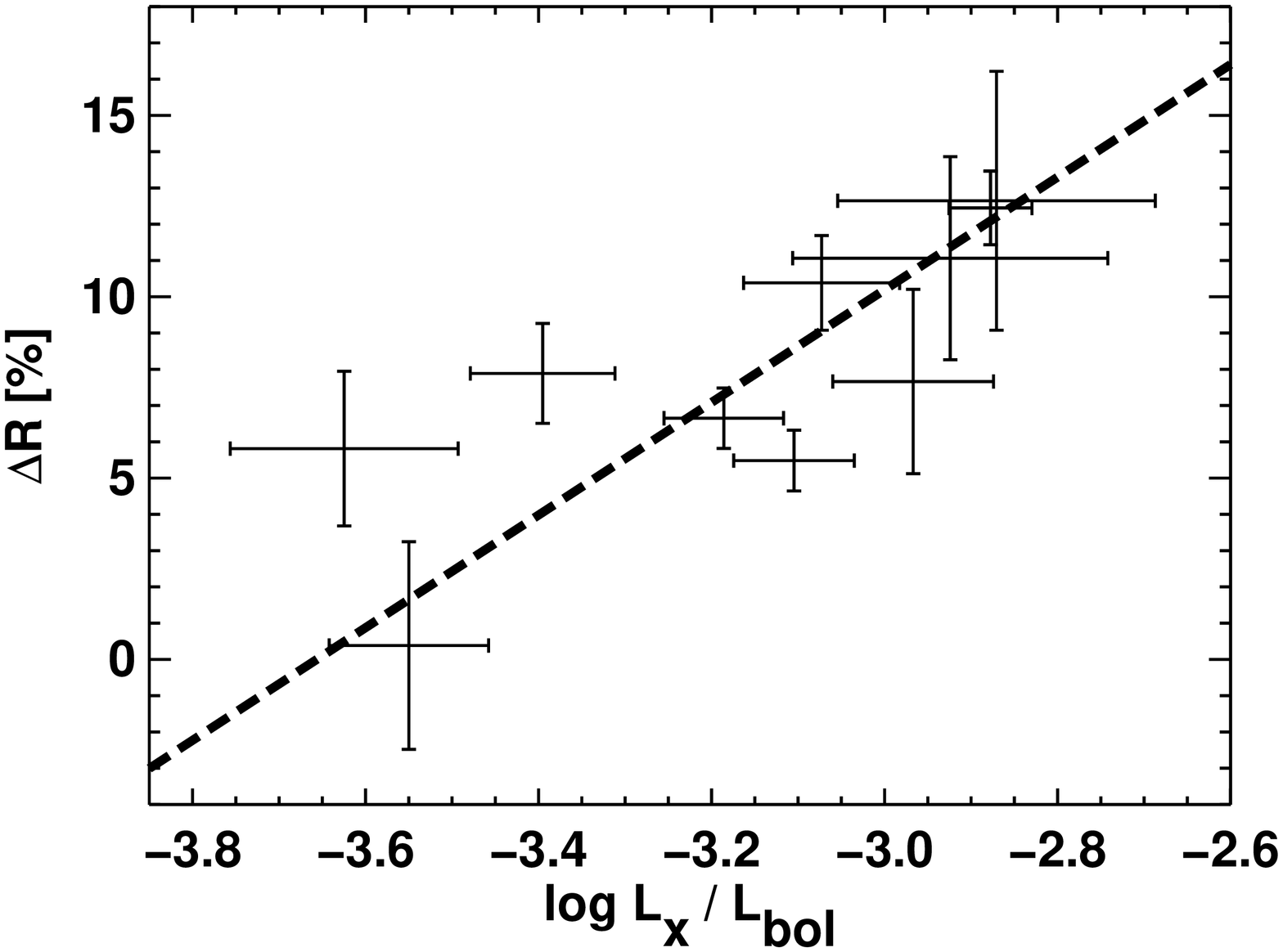}
\label{fig:lopezdata}
}
\subfigure[]{
%\caption{ 
%This figure is shown in color in the electronic version only.}
%\end{figure}

%\begin{figure}
%\epsscale{0.5}
\includegraphics[width=2.5in,angle=90]{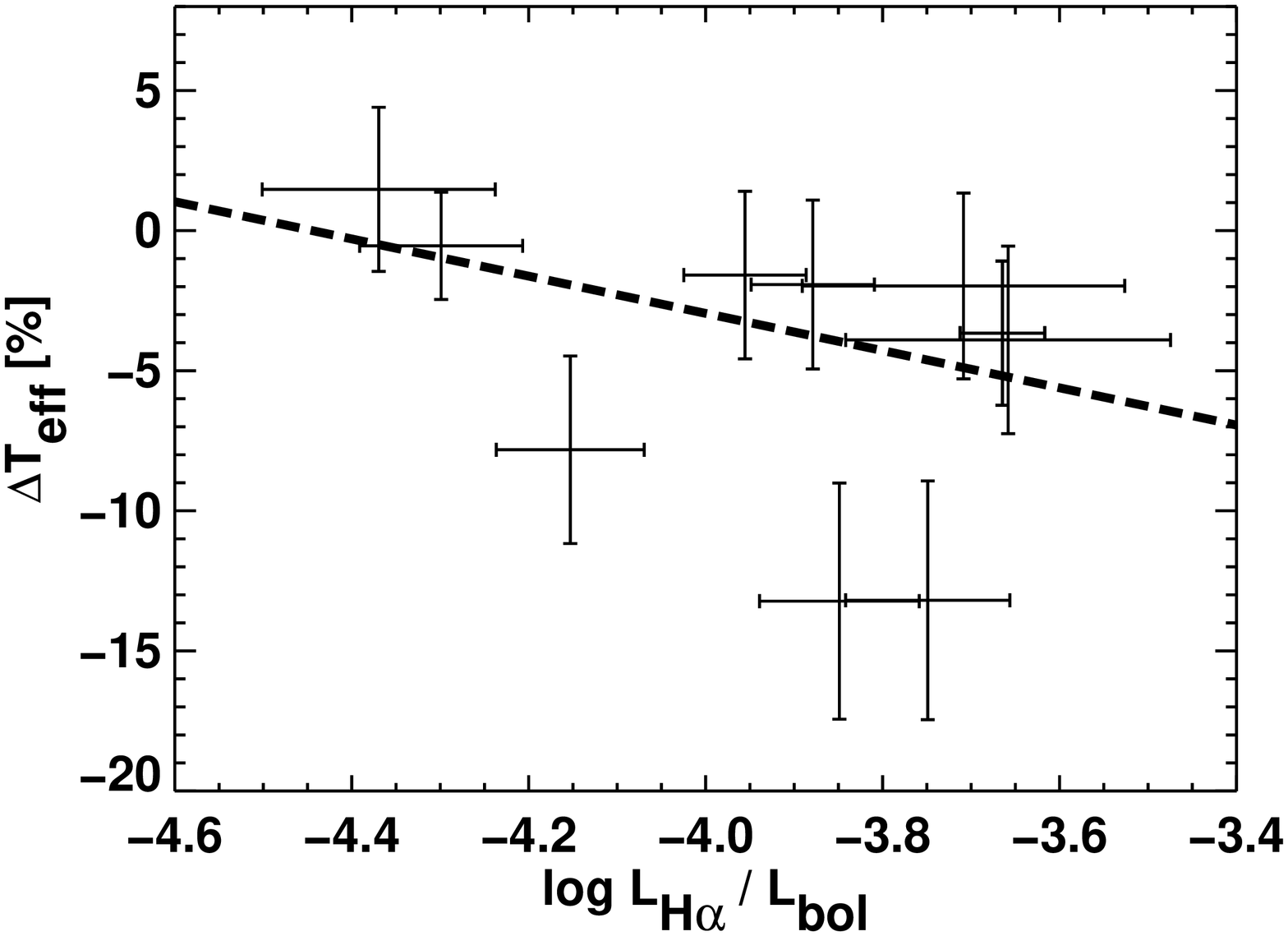}
\includegraphics[width=2.5in,angle=90]{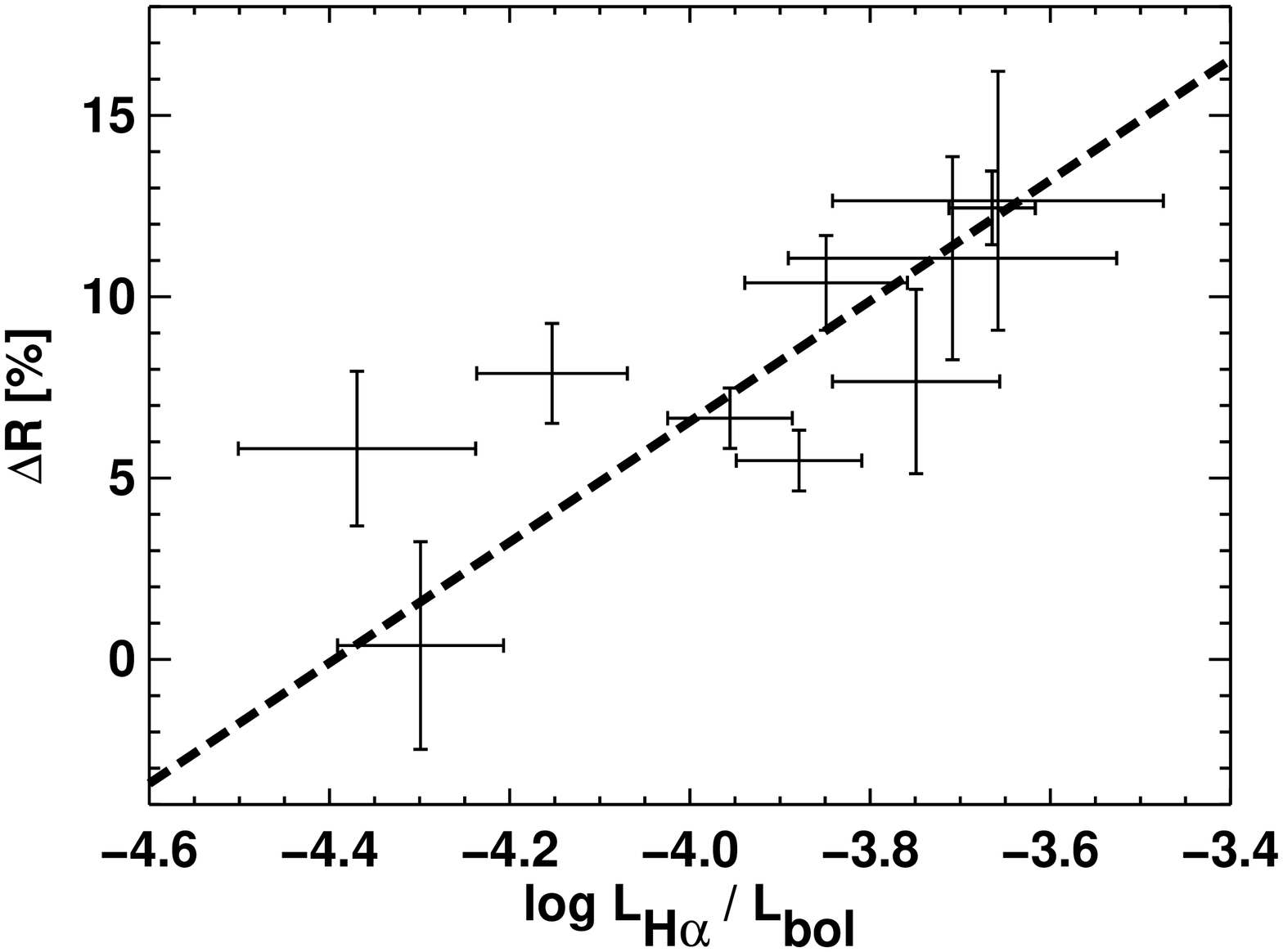}
\label{fig:lopezrelations}
}
\caption{Suppression of effective temperatures (left) and inflation of radii (right) for
low-mass eclipsing binaries from \citet{lopez07} as a function of
fractional X-ray luminosity (a) and as a function of fractional \ha\ luminosity (b). 
The $\Delta$\teff\ and $\Delta R$ are measured relative to the 3 Gyr isochrone of \citet{baraffe98}.
The fractional \ha\ luminosities are based on the fractional X-ray luminosities, using the
empirical X-ray-to-\ha\ luminosity relation in Fig.~\ref{fig:lxlh}. 
In all panels, the dashed line is a linear fit to the data. 
}
\end{figure}

We begin with the correlation of $\Delta R$ vs.\ \lx/\lbol\
already demonstrated in that work, which we rederived (Fig.~\ref{fig:lopezdata}) using the fundamental 
stellar data compiled in \citet{lopez07} and the same 3~Gyr isochrone of \citet{baraffe98}
as above. \citet{lopez07} did not discuss the complementary correlation with
$\Delta$\teff, but this information is also contained in the EB data,
so in Fig.~\ref{fig:lopezdata} we also derive the the relationship $\Delta$\teff\ vs.\ \lx/\lbol, 
again using the data compiled in \citet{lopez07} and the 3~Gyr isochrone of \citet{baraffe98}.
Linear fits are shown for both relations: 
$$ \Delta R/R\, [\%] = (15.5 \pm 2.9) \times \log L_X/L_{\rm bol} + (57 \pm 9) $$
$$ \Delta T_{\rm eff}/T_{\rm eff}\, [\%] = (-6.2 \pm 3.2) \times \log L_X/L_{\rm bol} + (-23 \pm 10) .$$

A Kendall's $\tau$ correlation test gives a 
null-hypothesis probability of 0.2\% for the correlation with $\Delta R$, indicating a
significant correlation at 99.8\% confidence \citep[see also][]{lopez07}. 
The correlation with $\Delta$\teff\ is not as strong, 
but is nonetheless modestly significant at 94\% confidence according to the Kendall's $\tau$ test.
Interestingly, the $\Delta R$ and $\Delta$\teff\ relations, which for the EB sample are measured
independently, very nearly offset one another in terms of their effect on \lbol.
For example, at $\log$~\lx/\lbol\ = $-3$, the relations give $\Delta R \approx$ 11\% and
$\Delta$\teff\ $\approx -5$\%, implying $\Delta$\lbol\ $\approx 1$\%. 
Thus \lbol\ is an approximately conserved quantity. 
%{\bf [need to return to this point in discussion]}. 

\begin{figure}
%\epsscale{0.5}
\includegraphics[width=4in,angle=-90]{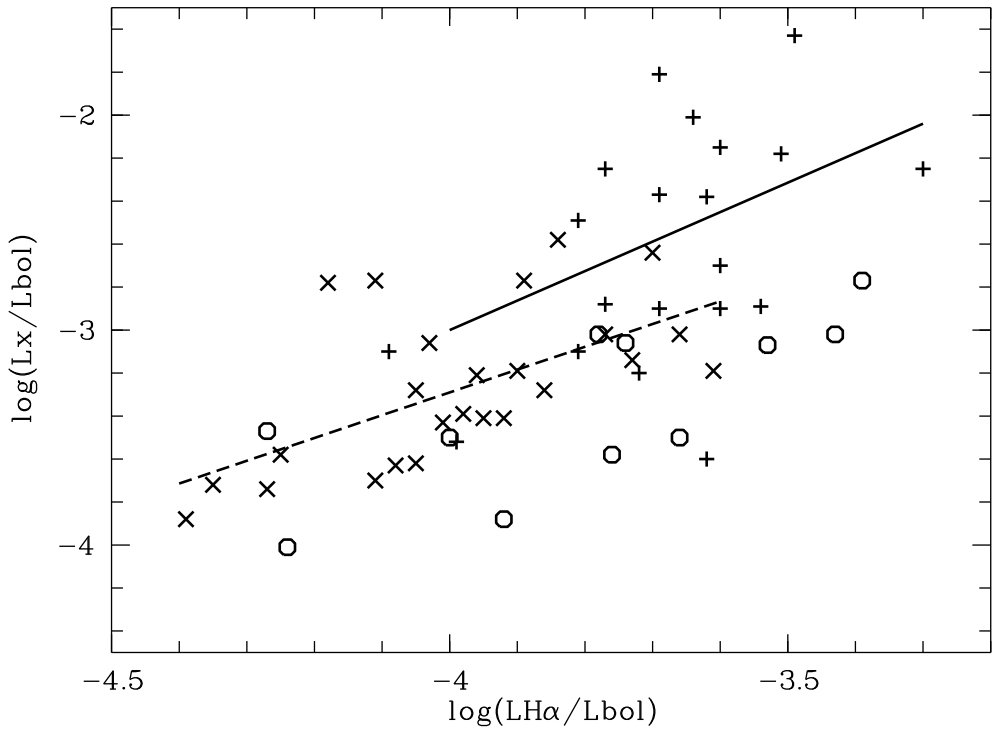}
\caption{\label{fig:lxlh}
Fractional X-ray luminosity vs.\ fractional \ha\ luminosity for low-mass star samples from
field dwarfs \citep[][crosses]{delfosse98}, young stars \citet[][plus symbols]{scholz07}, 
and field M-dwarfs \citep[][circles]{Reiners:2007}.
Linear fits to the low-mass star samples are shown as solid (fit to plus symbols)
and dashed (fit to crosses) lines.} 
%This figure is shown in color in the electronic version only.}
\end{figure}

To convert the observed \lx/\lbol\ into \lha/\lbol\ for use in Equations~\ref{eq:Trelation}--\ref{eq:Rrelation},
we use published activity data for two different samples having both X-ray and \ha\ emission
measurements (see Fig.~\ref{fig:lxlh}): low-mass stars in young associations 
\citep{scholz07} 
%(Scholz et al. 2007, plusses) 
and active M dwarfs in the field 
\citep{delfosse98}.
%(Delfosse et al. 1998, crosses). 
Fig.~\ref{fig:lxlh} demonstrates that there is a robust correlation between 
these two quantities over a wide range of activity levels, albeit with considerable 
scatter. A linear least-squares fit to this correlation is overplotted as solid and 
dashed line for the first and second sample, respectively. The activity data for M-type 
field dwarfs published by \citet{Reiners:2007} shows a similar trend 
(circles in Fig.~\ref{fig:lxlh}). We note that these correlations can be extended 
to lower activity levels, as they are consistent with the simultaneous activity data for 
BDs published by \citet[][their Table~5]{berger10}. 
%Berger et al. (2010, their Table 5). 
%The fit results are as follows:
For the young low-mass stars sample, the fit is: 
$$ \log (L_X/L_{\rm bol}) = (1.5 \pm 0.7) \times \log L_{\rm H \alpha}/L_{\rm bol} + (3.0 \pm 2.4) .$$
For the field M dwarfs and BDs, the fit is:
$$ \log (L_X/L_{\rm bol}) = (1.1 \pm 0.3) \times \log L_{\rm H \alpha}/L_{\rm bol} + (1.0 \pm 1.1) .$$
For our analysis we have chosen to use the second relation 
because it better reflects the range of X-ray activity levels observed in the 
\citet{lopez07} sample of low-mass EBs (see Fig.~\ref{fig:lopezdata}).
This is also a more conservative choice, in that it associates a lower
\lx/\lbol\ for a given \lha/\lbol, and therefore will predict smaller 
absolute $\Delta$\teff\ and $\Delta R$ offsets for a given observed \lha/\lbol.

Fig.~\ref{fig:lopezrelations} shows the same data as in Fig.~\ref{fig:lopezdata}, but now with
\lx/\lbol\ converted to \lha/\lbol. 
The resulting relations between the \teff\ and $R$ offsets vs.\ \lha/\lbol\ are 
discussed in Sec.~\ref{results}.

\section{Results \label{results}}

\subsection{Empirical Relations Linking \teff\ Suppression and Radius Inflation 
to \ha\ Emission\label{sec:relations}}
We have determined empirical relationships linking the degree of temperature suppression,
$\Delta$\teff/\teff, and radius inflation, $\Delta R/R$, to the fractional \ha\ luminosity, \lha/\lbol, 
using two independent samples and methods (see Figs.\ \ref{fig:pmsurelations} and \ref{fig:lopezrelations}). 
The fit coefficients from Eqs.~\ref{eq:Trelation}--\ref{eq:Rrelation} for the 
two samples are summarized in Table~\ref{tab:relations}. The fit coefficients are defined such that the 
left-hand side (LHS) of Eqs.~\ref{eq:Trelation}--\ref{eq:Rrelation} are in percent units. 

\begin{deluxetable}{lrrr}
\tablewidth{0pt}
\tablecolumns{4}
\tablehead{\colhead{} & \colhead{$m$} & \colhead{$b$} & \colhead{$\tau$ conf.\tablenotemark{a}}}
\tablecaption{\label{tab:relations}
Fit parameters for Equations \ref{eq:Trelation} and \ref{eq:Rrelation} }
\startdata
\cutinhead{Field M-dwarfs\tablenotemark{b}}
$\Delta$\teff &  $-3.12 \pm 3.15$  &  $-5.1 \pm 0.7$ & 92.9\% \\
$\Delta R$ &  $8.00 \pm 7.63$   &  $11.2 \pm 1.6$ & 91.2\% \\
\cutinhead{Eclipsing Binaries\tablenotemark{c}}
$\Delta$\teff &  $-6.64 \pm 3.47$  & $-3.0 \pm 1.0$ & 94.3\% \\
$\Delta R$ &  $16.64 \pm 3.15$  & $6.6 \pm 0.6$ & 99.8\% \\
\cutinhead{Averaged Final Relation\tablenotemark{d}}
$\Delta$\teff &  $-4.71 \pm 2.33$ & $-4.4 \pm 0.6$ & 96.0\% \\
$\Delta R$ &  $15.37 \pm 2.91$  & $7.1 \pm 0.6$ & 98.6\% \\
\enddata
\tablenotetext{a}{Statistical confidence of correlation from Kendall's $\tau$ test.}
\tablenotetext{b}{See Fig.~\ref{fig:pmsurelations}.}
\tablenotetext{c}{See Fig.~\ref{fig:lopezrelations}.}
\tablenotetext{d}{See Fig.~\ref{fig:finalrelations}.}
\tablecomments{The fit parameters are defined such that the LHS of 
Equations~\ref{eq:Trelation}--\ref{eq:Rrelation} are in percent units.} 
\end{deluxetable}

%\begin{deluxetable}{lrrr}
%\tablewidth{0pt}
%\tablecolumns{4}
%\tablehead{\colhead{} & \colhead{$m_R$} & \colhead{$b_R$} & \colhead{$\tau$ Conf.}}
%\tablecaption{\label{tab:Rrelations}
%Fit parameters for Equation \ref{eq:Rrelation}}
%\startdata
%Field M-dwarfs &  $-3.00 \pm 3.15$   &  $-17.04 \pm 12.68 $ \\
%Eclipsing binaries &  $-8.11 \pm 4.15$  & $-34.54 \pm 13.84$ \\
%\enddata
%\tablecomments{The fit parameters are defined such that the LHS of Equation~\ref{eq:Rrelation}
%is in percent units.} 
%\end{deluxetable}

The $\Delta$\teff\ and $\Delta R$ fit coefficients in Table~\ref{tab:relations}
are consistent to within 1--2$\sigma$ between the two samples,
although the formal uncertainties on the fit parameters are large for the field M-dwarf sample.
This likely reflects the large scatter in the \ha\ measurements for that sample, perhaps 
stemming from intrinsic stellar variability. In addition, for the field M-dwarf sample 
the determination of \lha/\lbol\ involved several calculated quantities (i.e., $R$, \lbol)
whereas for the EB sample these stellar parameters are measured directly and accurately. 
Indeed, for the field M-dwarfs the statistical significance of the $\Delta$\teff\ and $\Delta R$ correlations 
was stronger {\it before} we converted the directly measured EW(\ha) to \lha/\lbol. 
Even so, both the $\Delta$\teff\ and the $\Delta R$ relations for both samples are 
confirmed to be significantly correlated with \lha/\lbol\ 
%at greater than 95\% statistical significance 
according to a Kendall's $\tau$ test (Table~\ref{tab:relations}). Being a non-parametric
rank-correlation test, Kendall's $\tau$ does not depend on the assumed functional form of the
relationship, and thus robustly indicates the presence of a correlation even if the significance
of the assumed functional parameters is modest. 

\begin{figure}
%\epsscale{0.5}
\includegraphics[width=2.5in,angle=90]{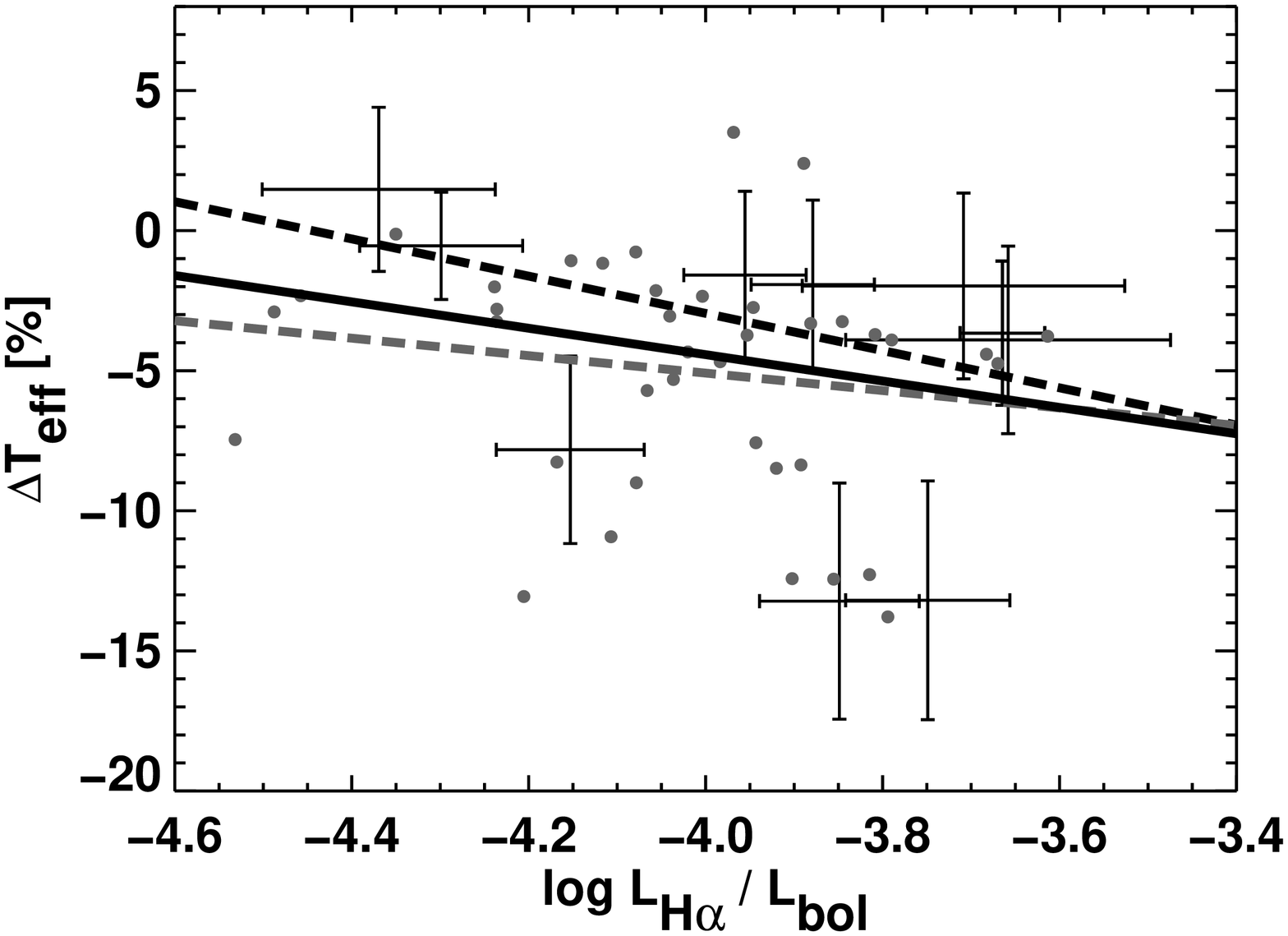}
\includegraphics[width=2.5in,angle=90]{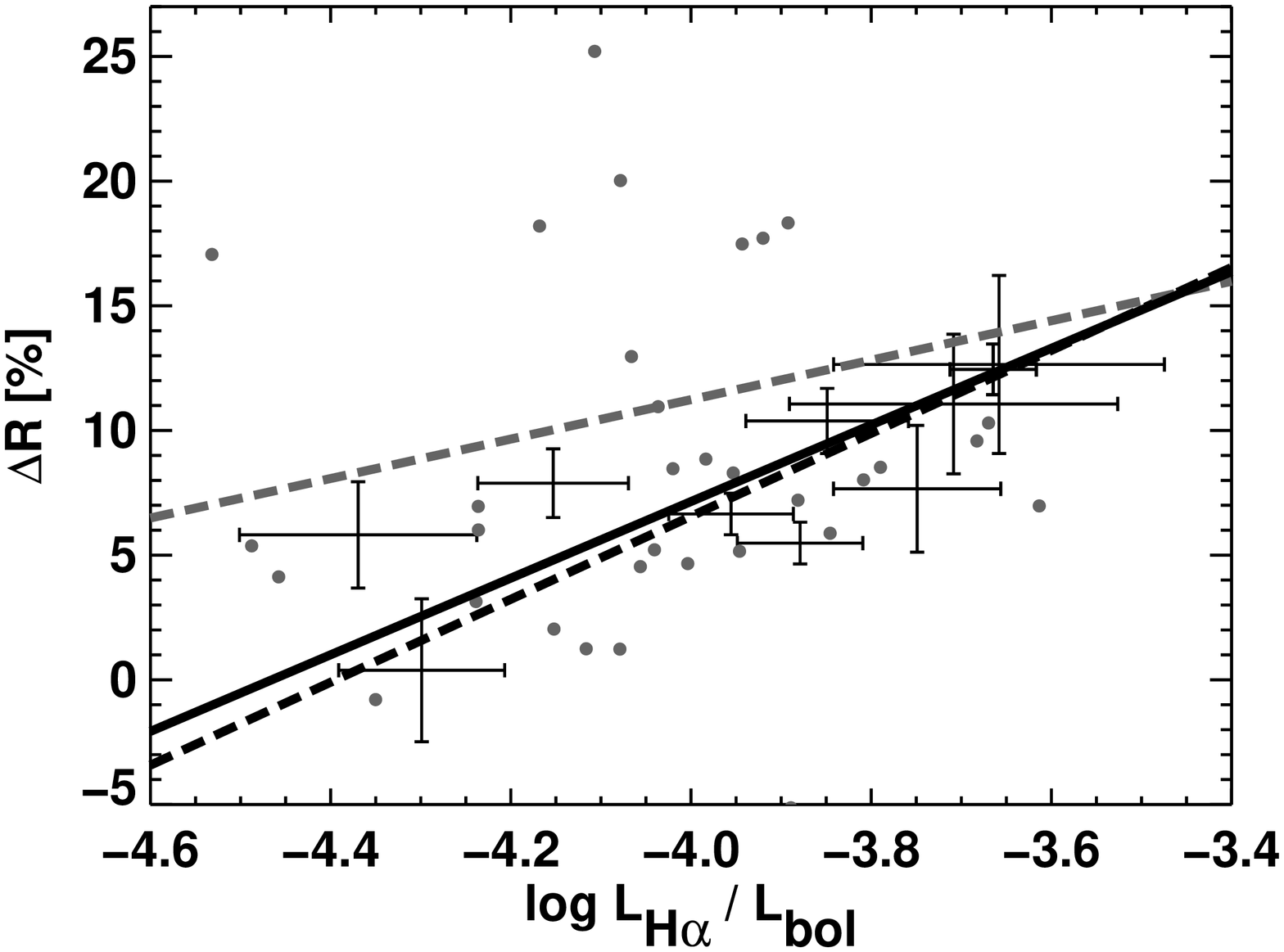}
\caption{\label{fig:finalrelations} 
Temperature suppression (left) and radius inflation (right) as a function of fractional \ha\
luminosity for both field M-dwarfs (filled symbols) and low-mass eclipsing binaries (error bars). 
Relations from Figs.~\ref{fig:pmsurelations}
and \ref{fig:lopezrelations} are reproduced (dashed lines). The final averaged best-fit relation
in each panel is shown as a solid line.
See Table~\ref{tab:relations} for the linear fit coefficients to these relations
(Eqs.~\ref{eq:Trelation} and \ref{eq:Rrelation}).
}
\end{figure}

For our final best-fit relation, we calculated the weighted average of the fit parameters
for the two samples.
The resulting best-fit coefficients for Eqs.~\ref{eq:Trelation} and \ref{eq:Rrelation} 
are also listed in Table~\ref{tab:relations}, and the corresponding final fits shown
in Figure~\ref{fig:finalrelations}. 
As is evident in Fig.~\ref{fig:finalrelations}, the r.m.s.\ scatter of the data about the mean relations is large
(4.2\% and 9.2\% for $\Delta$\teff\ and $\Delta R$, respectively),
driven primarily by the scatter in the field M-dwarf data (see Sec.~\ref{mdwarfs}).
Nonetheless, the statistical significance of the final combined relations is strong (Table~\ref{tab:relations}). 
%of $m=-8.11 \pm 4.15$ and $b=-34.54 \pm 13.84$, defined such that the LHS of Equation~\ref{eq:Trelation} is in percent units. 
%{\bf Can we say why we have chosen to use the second relation as the 'best fit'?}
%The two relations are consistent with one another. 
%{\bf [Note: some discussion on the statistical significance of the correlations and their consistency is needed.]}

The input samples used to determine these relations include low-mass stars with $0.2<M<1$ \msun. 
There is no evidence that the \teff\ suppression and $R$ inflation effects change qualitatively at
lower masses (see Fig.~\ref{fig:pmsudata}), and we are able to validate them successfully at BD masses
(see below). However, the relations appear to diverge at masses $\gtrsim$0.8 \msun\ (see Fig.~\ref{fig:pmsudata})
and so we caution against their use at such high masses.
Given the range of \lha/\lbol\ spanned by the input data samples, these relations are applicable for
$-4.6 \lesssim \log$~\lha/\lbol\ $\lesssim -3.4$ (see Fig.~\ref{fig:finalrelations}), the upper limit
corresponding approximately to the empirical ``saturation" level of $\log$~\lha/\lbol\ $\approx -3.3$
observed in low-mass stars and BDs \citep[e.g.][]{barrado03}. 
Therefore the relations likely cannot be extrapolated to $\log$~\lha/\lbol\ $> -3.3$. Note also that
for $\log$~\lha/\lbol\ $\lesssim -4.6$, Eqs.~\ref{eq:Trelation}--\ref{eq:Rrelation} give {\it positive}
$\Delta$\teff\ and {\it negative} $\Delta R$, which is likely not physical. More likely the $\Delta$\teff\
and $\Delta R$ offsets simply approach zero at very low activity levels, and thus we caution that
our relations should not be extrapolated to arbitrarily low \lha.

\subsection{Application to \2m \label{2m0535}}

%{\bf [Trent's note:
%I was thinking about adding an error bar to the DeltaTeff/Teff number, but
%it turns out that it would be 7\% +- 12\% (radius is 14\% +- 15\%).  That
%obviously wouldn't look very good!  Part of this could simply be
%related to the fact that the coefficient error bars aren't necessarily
%reflective of reality.  Since the points used in at least one of the
%fits did not have error bars (not sure if you did a weighted fit for
%the smaller sample), the errors just come from some Numerical Recipes
%formula that I think essentially tries to make chi2 $\sim$ 1.  Therefore,
%I wonder if we should even bother quoting error bars on the fit
%coefficients?  This inoculates us against people who might get hung up
%on the large errors, and instead we could include a discussion of how
%variability in Halpha results in somewhat uncertain Teff corrections
%(a much more modest effect).]} 

%{\bf [Keivan will calculate proper errors on the \2m\ corrections using covariances,
%and also calculate rms about fits. Trent will also think about this.]} 

%We apply the above relation to the BD EB \2m, 
\2m\ is to date the only known system of BDs with directly and accurately measured masses, radii, \teff, and \lha,
all at a well constrained system age \citep{stassun06,stassun07,gomez09}.
\2m\ is therefore an important empirical test case for assessing the efficacy 
of the activity-based $\Delta$\teff\ and $\Delta R$ relations that we have determined above.
The known masses of the primary and secondary BDs in \2m\ are 60$\pm$3 and 39$\pm$2 \mjup, respectively. 

\begin{figure}
%\epsscale{0.75}
\includegraphics[scale=0.8,angle=90]{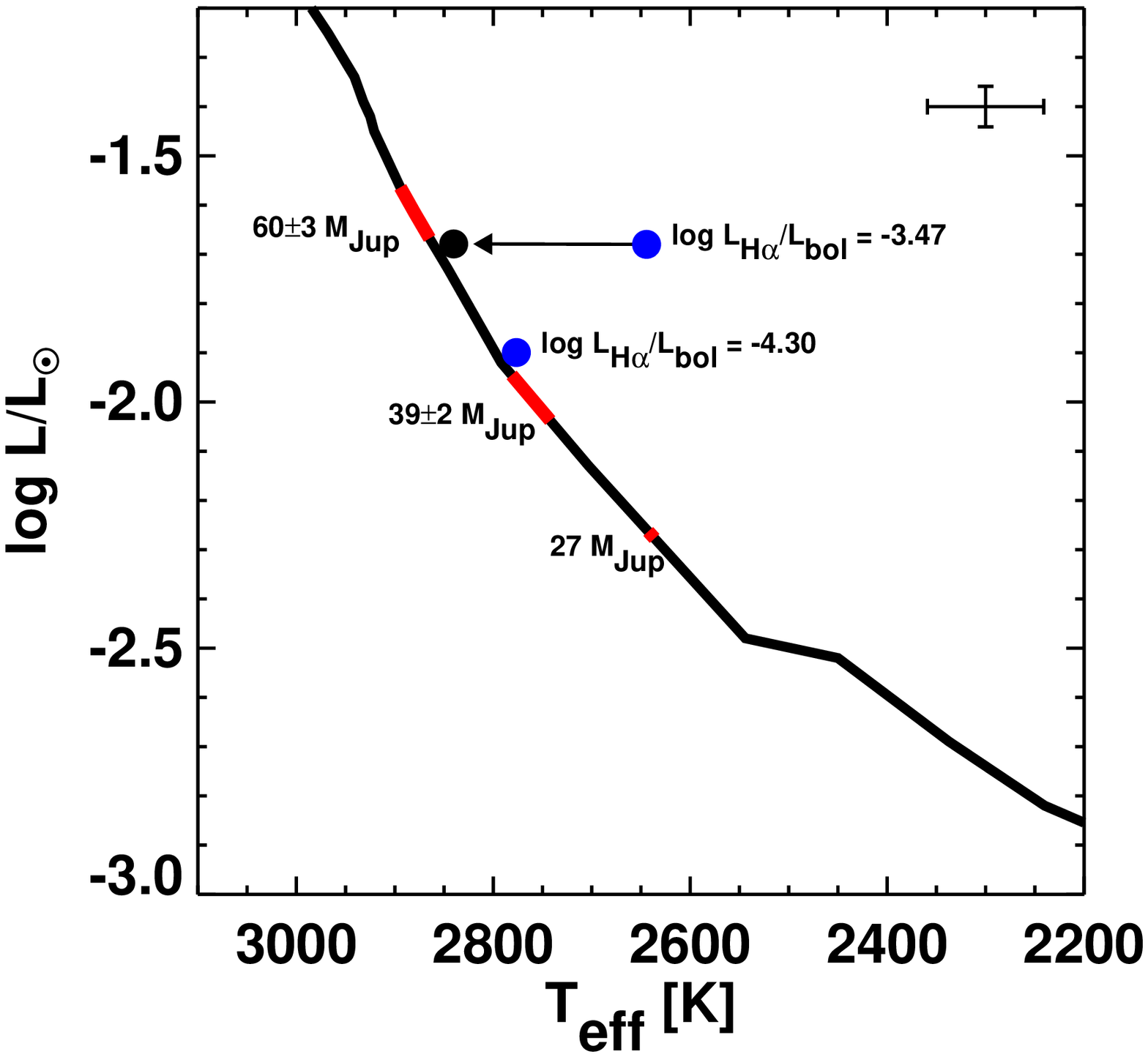}
\caption{\label{fig:2m0535} 
Hertszprung-Russell diagram for the primary and secondary components of the brown-dwarf
eclipsing binary \2m\ in the $\sim$1-Myr Orion Nebula Cluster \citep{stassun06,stassun07,gomez09}. 
The measured \teff\ and \lbol\ (the latter calculated from the directly measured \teff's and 
radii; see Sec.~\ref{2m0535}) for both brown dwarfs are represented as blue symbols. 
Measurement uncertainties in \teff\ and \lbol\ are represented by the error bars at upper right.
The dynamically measured masses of the primary and secondary are 60$\pm$3 and 39$\pm$2 \mjup,
respectively, represented as red bars on the 1-Myr theoretical isochrone of \citet{baraffe98}.
The measured \lha/\lbol\ for the two components from \citet{reiners07} are indicated next to the
blue symbols. The inactive secondary appears close to its expected position on the isochrone,
whereas the active primary appears far cooler than expected. The primary therefore
appears to be much younger than the secondary and to have a mass of only $\approx$27 \mjup\ 
based on its observed \teff, a factor of 2 lower than its true mass. 
Shifting the position of the active primary (arrow) using our empirically calibrated \ha-based
relations for \teff\ suppression and radius inflation brings the primary into much closer
agreement with its theoretically expected position in the HR diagram (black symbol);
this is where the active primary would be if it were not active.
This figure is shown in color in the electronic version only.
}
\end{figure}

To demonstrate how the {\it apparent} masses of the \2m\ primary and secondary BDs are altered by our
empirical relations, we calculate how the observed positions of the two BDs in the H-R diagram are altered
by adjusting the observed \teff\ and $R$ using Eqs.~\ref{eq:Trelation}--\ref{eq:Rrelation} and
Table~\ref{tab:relations}. 
In effect, we are using our empirical relations to show how the \2m\ system would appear in the H-R
diagram were the system completely inactive.
The spectral type of M6.5$\pm$0.5 for \2m\ determined by \citet{stassun06} from high-resolution $H$-band 
spectroscopy implies an average \teff=2688$\pm$55~K, based on the near-infrared spectral-type--\teff\ relation
of \citet{slesnick04}. Weighting this average \teff\ by the $H$-band primary-to-secondary flux ratio of 1.6:1
found by \citet{stassun07} and using the accurately determined \teff\ ratio of 
$T_1 / T_2 = 0.952 \pm 0.004$ \citep{gomez09}
then gives \teff=2640$\pm$60 and \teff=2770$\pm$60 K for the primary and secondary components, respectively.
These individual \teff\ together with the accurately measured individual radii \citep{stassun07,gomez09}
then give the individual \lbol.
These observed \teff\ and \lbol\ for the primary and secondary components are represented in Fig.~\ref{fig:2m0535} 
as blue symbols, from which one would infer masses (using the Baraffe et al.\ evolutionary models) of 
$\approx$27 and $\approx$40 \mjup, respectively, based on the observed \teff. That is, the inactive secondary
is inferred to have approximately its true mass, but the active primary
appears to have a mass that is a factor of 2 lower than its true mass,
and moreover the primary appears to be much younger than its (presumably coeval) companion. 
%\textbf{Keivan: Should we mention here that at this \teff\ and \lbol\ the primary actually falls completely off of the model isochrones?  (This is also a problem for field brown dwarfs that just live in no man's land on the H-R diagram -- how do you infer model properties? -- at least here there is the possibility of an age difference or something.)}
%{\bf [... without applying a correction one would infer non-coeval system and an incorrect mass.]} 

We use the \lha/\lbol\ measurements of \citet{reiners07}, who found 
$\log$~\lha/\lbol\ = $-3.47$ for the active primary and 
$\log$~\lha/\lbol\ $< -4.30$ (upper limit) for the inactive secondary.
From Eqs.~\ref{eq:Trelation} and \ref{eq:Rrelation},
we find that the primary has been displaced from its inactive \teff\ by an amount
%\sout{the resulting $\Delta$\teff\ for the primary is} 
$\Delta$\teff\ = $-6.9\pm1.4$\%, and 
displaced from its inactive $R$ by an amount $\Delta R = 15.2\pm1.7$\%.
%(implying a very nearly constant \lbol). 
For the secondary, the resulting $\Delta$\teff\ is at most $-3.0\pm0.9$\%, and $\Delta R$ is at most $2.5\pm1.1$\%.
%(again implying a nearly constant \lbol).
For both BDs the $\Delta$\teff\ and $\Delta R$ 
displacements nearly cancel such that \lbol\ is preserved to within 1--3\%. 

As shown in Fig.~\ref{fig:2m0535}, shifting the H-R diagram position of the 
primary BD according to its calculated $\Delta$\teff\ and $\Delta R$ %\sout{puts it into}
shows that, if it were inactive, it would be in
remarkably good agreement with the theoretically expected position for its known mass.
The observed \lha/\lbol\ of the active primary is near the ``saturation" limit and therefore
near the upper limit of observed activity levels in young low-mass objects, for which the
\ha\ emission is found to be mostly non-variable \citep[e.g.][]{bell12}. Thus we do not
expect a large contribution to the uncertainty of the active primary in Fig.~\ref{fig:2m0535}
due to \ha\ variability. 
We do not shift the position of the secondary as its \lha\ is an upper limit only. 
%and the shifted position assuming the maximum correction is comparable to the plotting symbol size in the figure. 

\section{Discussion\label{summary}}

%{\bf [What are some additional concerns/criticisms that we should anticipate and therefore address here?]}

\subsection{\ha, magnetic activity, and cooled/inflated stars}

In developing these empirical relationships, we posit a correlation between chromospheric activity, radius inflation, 
and \teff\ suppression, where activity is implicitly tied to magnetic fields. \citet{chabrier07} and \citet{macdonald2009} have suggested that a sufficiently strong field could suppress convection, inhibit heat transfer, and thus inflate (and cool) the stellar surface. Since such a field  would also likely result in chromospheric activity, one might therefore expect the correlations that we have derived.  However, there are two significant outstanding questions in creating the above link: (1) Is the relation between \ha\ emission, field strength, and radius inflation / \teff\ suppression monotonic? (2) Are the magnetic fields in fully convective objects strong enough to account for the radius inflation and \teff\ suppression?

First, it remains unclear how well \ha\ emission correlates with the strength of the magnetic field.  \citet{Reiners:2007} have attempted to relate field strength and \lha/\lbol\ using the FeH line. They find a correlation between the two, but with different power law scalings for different spectral types. Whether this is an age or mass effect remains uncertain. While this correlation is encouraging for the empirical relations we provide here, \citet{Reiners:2007} use slowly rotating field stars, whose magnetic properties may differ from the most active (younger) objects in which we are interested.
As noted previously, \ha\ emission is also variable, though  the most active objects typically have the lowest variability \citep{bell12}. Concurrent multi-wavelength monitoring of multiple chromospheric tracers including radio and X-ray, show that the individual tracers do not always vary on the same timescales \citep{Berger:2006,Berger:2008}. Thus it is unclear that any individual wavelength range is representative of the total energy contained in the field. 
Despite these different trends in variability, that \lx\ also correlates with \teff\ suppression and radius inflation (see Fig.~\ref{fig:lopezdata}) corroborates our use of chromospheric activity to account for the displacement of objects from model isochrones \citep[see also, e.g.,][and references therein]{morales08}.
%\sout{[Kaitlin: I think that last sentence needs to be reworded? I think you just want to say that the fact
%that \teff\ suppression and $R$ inflation *both* correlate with activity gives credence to both correlations?
%Should we say that the preservation of \lbol\ further corroborates the relations?]}

The second concern, whether or not magnetic fields are sufficiently powerful to inflate the stars 
and suppress their surface temperatures, is currently impossible to address without appealing to models. \citet{Browning:2008} has shown in global numerical models that fully convective stars can host large Kilogauss strength fields, but these alone are too weak to produce the observed radius inflation and \teff\ suppression in \2m\ \citep{macdonald2009}. \citet{chabrier07} also suggest that fully convective objects should be less affected by the same convective inefficiencies invoked to explain radius discrepancies at higher masses.  It is possible that a combination of rotation and magnetic activity contribute to both inflation/suppression and \ha\ emission in such a way as to produce our empirical relation without a causal correspondence between \ha\ and the magnetic field. 

An alternative explanation for stars with \teff\ deficits is a spot covered surface \citep[e.g.][]{Lacy1977}. For the case of \2m, \citet{mohanty10} have argued that a model for the active primary with 70\% (axisymmetric) spot coverage could explain the peculiar mass-temperature relationship in that system. Since spot coverage is also controlled by magnetic fields, a correlation with \ha\ might still exist. It remains unclear how to interpret spot coverage of greater than 50\%; perhaps the analogy with solar type spots breaks down in such extreme systems. 
%{\bf [KGS to add sentence re Monhanty 2012 result.]} 
Indeed, \citet{mohanty12} have now shown from a spectral fine-analysis of the \2m\ system, observed 
at high resolving power during both primary and secondary eclipses, that such a large-spot scenario is strongly
disfavored as an explanation for the \teff\ suppression of the primary BD in \2m. Thus, while the
\teff\ suppression mechanism produces a clear correlation with chromospheric \ha\ activity as we have shown
here, it evidently does not in all cases effect this correlation directly through surface spots.

%{\bf Note from Keivan: Can we include some discussion of the implications Mohanty et al (2010) who showed that
%the Teff-suppressed spectrum of the primary BD in 2M0535 is probably *not* due to spots, and so this argues for
%global Teff suppression?} 

\subsection{Impact on estimates of object masses} 

%{\bf [Trent's note: 
%The other (bigger) issue here is
%that there is no real discussion of how measurement errors figure into
%mass estimates.  For example, for a given (Teff,age) $\rightarrow$ mass, how do
%the typical errors in Teff and age propagate into a typical error in
%mass?  This would be useful to gauge how important the activity effect
%is in the grand scheme of things.  I could come up with something
%along these lines if I knew what typical Teff and age errors are.]
%[Aleks says: 
%Either I'm misunderstanding something, or this information is already buried somewhere
%in Sect. 4.2. For 1Myr the maximum error in mass is in the range of factor 2. For older
%objects it's less. This is when we start with Teff. When we start with luminosity, the main
%uncertainty would be the spread in the HR diagram along the y-axis, which is due to a
%number of effects, including age spread. One could look at some of the HR diagrams in the
%papers I've cited in this section (Da Rio, Luhman, Hillenbrand) and see how much that is,
%and then go from there.]
%[Aleks's note:
%Trent, I also think it might be interesting to add a paragraph pointing out the additional
%problems with the Teffs in this mass regime (opacities, dust clouds, etc.), just to be clear
%about the problem.]} 

%{\bf [Trent to write a paragraph incorporating the above points, somewhere down below.]} 

The vast majority of masses for stars and BDs in young clusters can only be determined by comparison with 
theoretical evolutionary tracks in H-R diagrams, using either luminosity or \teff\ or both 
\citep[e.g.][]{kh95,hill97,luhman98,moraux07,scholz12}. While direct mass measurements are available for calibration 
of these models at higher masses ($M \gtrsim$ 0.3 \msun; e.g., \citealp{hillenbrand04}), at the lowest stellar masses 
and into the substellar regime these models are essentially untested.  Therefore, if there is indeed a relation between 
magnetic activity and $R$ inflation / \teff\ suppression, this will also affect estimates of stellar and substellar 
masses derived from \teff, especially at young ages when activity levels are high. In the following, we examine 
two ways of estimating masses that are commonly used in the literature and investigate the impact of magnetic activity on the derived masses. 

\begin{figure}
%\epsscale{0.5}
\includegraphics[width=3.25in,angle=-90]{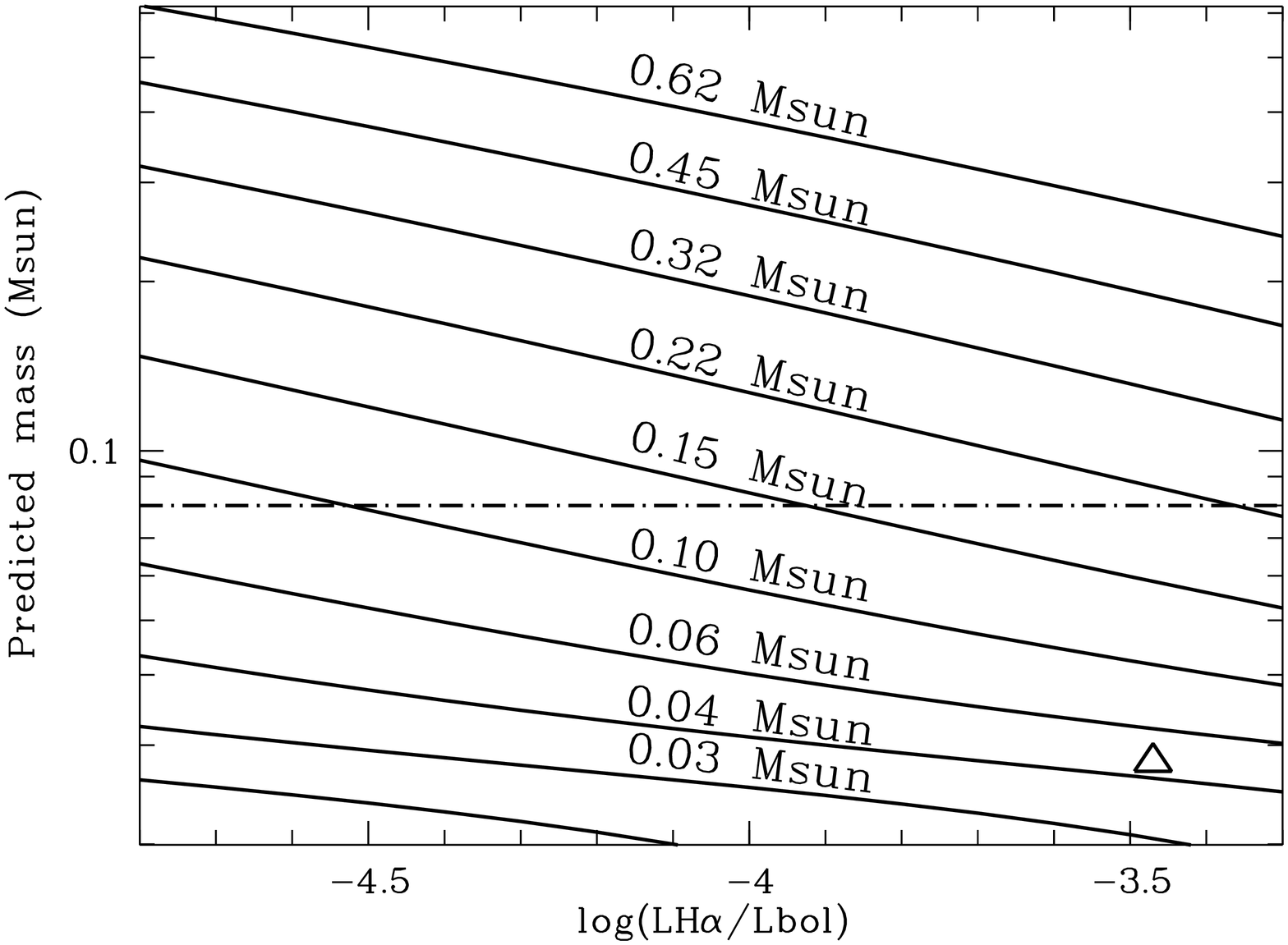} \\
\includegraphics[width=3.25in,angle=-90]{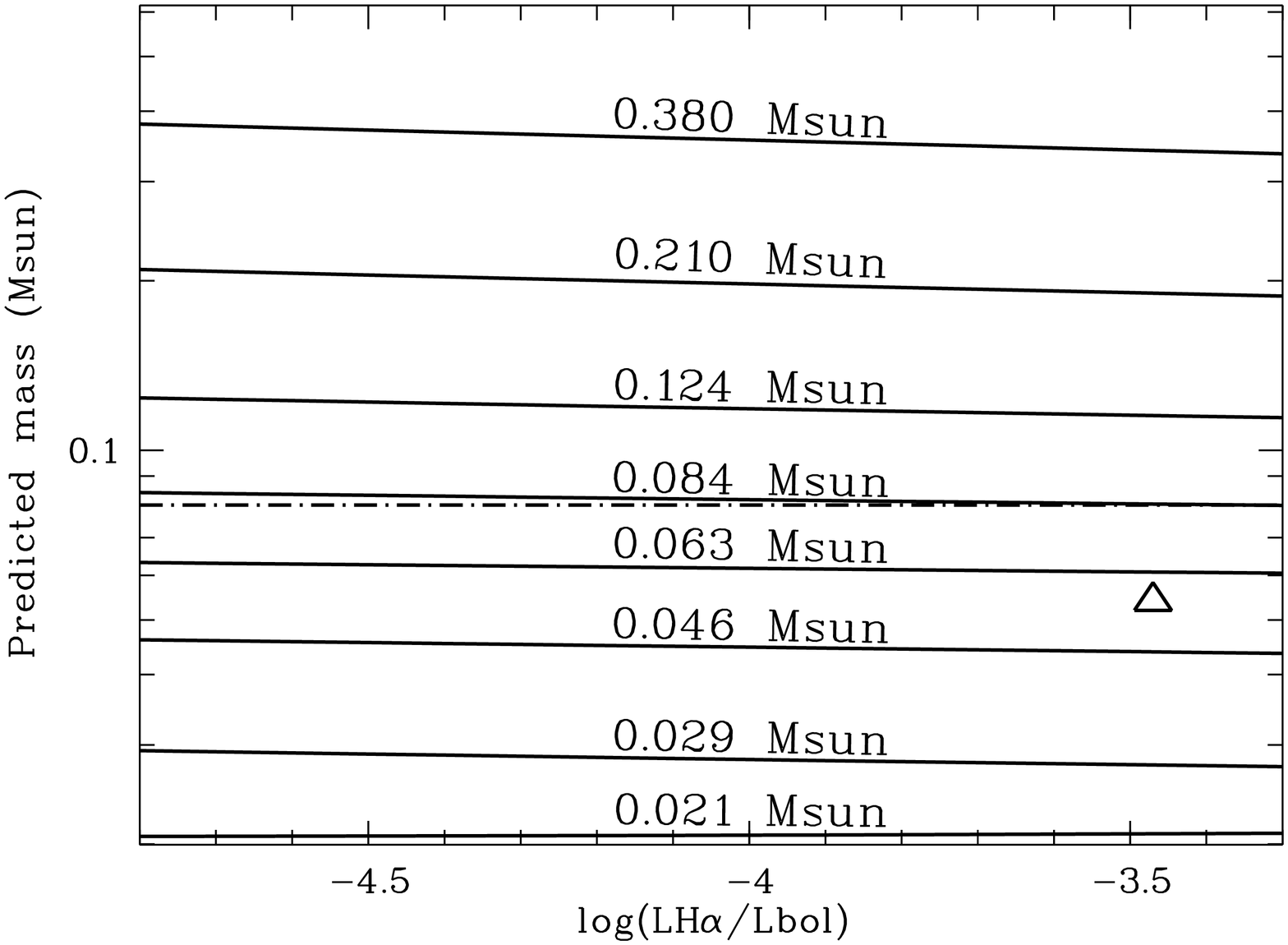} 
\caption{\label{fig:imf}
Inferred masses of low-mass stars and brown dwarfs as a function of \lha/\lbol, derived using the BCAH98 and DUSTY evolutionary 
tracks and the empirical relation between \teff\ suppression and magnetic activity derived in this paper. 
In each panel, the true object masses are labelled on the curves, whereas the masses that would be inferred at a given
activity level are shown on the ordinate. At low activity levels ($\log$~\lha/\lbol\ $\lesssim -4.6$), the inferred
mass is close to the true mass. 
Top panel: Masses are inferred directly from the observed \teff, which is susceptible to suppression at high
activity levels, and therefore the inferred masses are systematically lower than the true masses.
Bottom panel: Masses are inferred from the bolometric luminosity, estimated from the $K$-band absolute magnitude
and bolometric correction (as would be appropriate for an object of known distance). Here the \teff\ suppression
effect enters only weakly via the \teff-dependency of the $K$-band bolometric correction.
The dash-dotted line in each panel shows the substellar 
limit. The triangle in each panel shows the position of the primary component in the eclipsing binary brown dwarf \2m, 
which has a true mass of 0.06$\,M_{\odot}$.}
\end{figure}

First we use \teff\ to estimate masses from model isochrones at 1~Myr.  At each model mass, we consider a range of activity 
levels and apply offsets to the model \teff\ based on our Eq.~\ref{eq:Trelation} (using ``averaged final'' coefficients from 
Table~\ref{tab:relations}).  We then use these suppressed \teff\ values to estimate the masses that would be inferred from the isochrone.  The resulting curves are shown in the upper panel of Fig.~\ref{fig:imf}. (Note that for $\log$~\lha/\lbol\ = $-4.7$ the offset with Eq.~\ref{eq:Trelation} becomes zero.)  We consider masses of 0.02--0.60~\msun\ (i.e., \teff = 2500--3500~K), for which we use a combination of BCAH \citep{baraffe98} and Dusty \citep{allard01} isochrones.  (Using BCAH models at masses $\geq$0.1~\msun\ and Dusty models at lower masses, we fit a third order polynomial to mass as a function of \teff.)

As expected, this procedure leads to a systematic {\it underestimation} of the masses. At high levels of activity, the effect 
can be substantial. For $\log$~\lha/\lbol\ = $-3.3$, which corresponds to the saturation limit in young low-mass stars 
\citep{barrado03} and the upper limit of the observed \ha\ activity in young associations \citep{scholz07}, the mass estimates 
are a factor of $\sim 2$ lower than for objects with low levels of magnetic activity ($\log$~\lha/\lbol\ $<-4.5$). 
In Fig.~\ref{fig:imf} we
overplot the datapoint for the primary component of the eclipsing BD \2m. As discussed above, its mass 
would be estimated to be 0.03 \msun, whereas its true mass is about twice as high. 
%
%{\bf KMK: is this calculation redundant? -- AS: No, why? -- TJD: Kaitlin means that you have just computed the arrow going in the opposite direction from the %way it is shown in Fig. 1, so in that sense it is redundant.}
% now rephrased
We note that because the mass--\teff\ relationship is less steep at older ages, the underestimation of mass is less severe
at older ages ($\gtrsim$100~Myr).

In the second test case, we use \lbol\ derived from $K$-band absolute magnitudes to derive masses. In this case the influence of magnetic
activity is introduced by the \teff\ dependence of the bolometric correction, since the bolometric luminosities are (as outlined 
above) practically not affected.  We consider the same range of masses as in the first case, and we compute spectral types from
the suppressed values of \teff\ using the relation of \citet[][fitted linearly]{luhman03}.  We then computed $K$-band bolometric 
corrections from \citet{slesnick04} from the spectral types, and we combine these with the model $K$-band absolute 
magnitudes to find \lbol\ (we used a linear fit of model $M_K$ as a function of \teff).  Finally, we estimated masses from
the model isochrone and \lbol, and the result is shown in the lower panel of Fig.~\ref{fig:imf}.

This second method still leads to an underestimate of the masses, but the effect is much smaller than when directly estimating
masses from \teff. The curves are almost flat, i.e.\ the estimated masses depend insignificantly on the level of magnetic activity. The change of the bolometric corrections with increasing magnetic activity is quite small (at most 10\%), resulting in relatively minor changes of a few percent in the mass estimate. With this method, the mass for the primary component of the eclipsing BD \2m\ would be 0.054 \msun\ and thus only marginally smaller than the true value. 

%Note that our analysis uses evolutionary tracks which do not take into account the influence of magnetic fields. It is likely 
%that magnetic  fields that are sufficiently strong to suppress convection might manifest themselves in ways more complex than 
%enhanced chromospheric emission. Our results should be seen as a first-order approximation of the effect of activity on mass 
%determinations. 
%%In the future this type of analysis should be carried out with models including magnetic fields. 

%\textbf{NEW PARAGRAPH BY TRENT:} 
These effects are systematic biases in estimated masses due to \teff\ suppression, so this will always be significant for statistical studies of populations (e.g., in young clusters).  For individual objects, this bias may be somewhat lower than the uncertainty in \teff\ due simply to our limited ability to accurately model spectra of 2500--3500~K objects.  
%At higher temperatures ($>$3000~K), models have been tested against single, inactive stars with direct radius measurements from interferometry.  XXX find YYY.  At the lower temperatures direct radii are not yet available, but 
For example, the best benchmark objects currently available over the 2500--3000~K range---very low-mass stars and BDs with dynamical mass measurements---show that model atmospheres likely harbor large systematic errors ($\approx$250~K; \citealp{2010ApJ...721.1725D}).  This uncertainty corresponds to a fractional error in \teff\ of $\approx$8\%--10\%, which is somewhat larger than the typical offset we find due to \teff\ suppression.

An important application of \teff-based mass estimates is the determination of mass functions (MFs) for young clusters. Typical 
early to mid M-type dwarfs with ages from 10 to 100\,Myr exhibit \ha\ emission in the range $\log$~\lha/\lbol\ = $-4.2$ to $-3.3$
\citep{barrado03,scholz07}. For BDs at the same ages the available data are sparse but indicate an upper limit around 
$-3.7$ \citep{barrado03}. Thus, the upper panel of Fig.~\ref{fig:imf} implies that objects inferred to be BDs from
\teff-based (or spectral type based) mass estimates could actually be low-mass stars since masses will be underestimated by
up to a factor of 2.  

We further evaluated the impact of this effect on measurements of the slope of the mass function, $\alpha$ (in $dN/dM \propto M^{-\alpha}$).
For this purpose, we assume a measured slope of $\alpha = 0.6$, which is consistent with a number of studies in very young clusters
(see Scholz et al.\ 2012, submitted, and references therein), calculate the mass function based on that slope for $M<0.6\,M_{\odot}$, 
correct the masses for a given level of \ha\ emission using the results given above, and re-determine the slope $\alpha$. If magnetic 
activity and thus the level of \ha\ emission is constant across the low-mass regime, $\alpha$ is practically unchanged, because all 
masses will be underestimated by about the same factor. Based on the available measurements, however, it seems more realistic to assume 
that BDs have on average a lower level of activity than low-mass stars \citep{barrado03}. If we start with 
$\log$~\lha/\lbol\ = $-4.0$ for BDs and $-3.5$ for low-mass stars, the masses have to be corrected by factors of 1.3--1.7 for
BDs, and by 2.2--2.5 for low-mass stars. As a result, the slope changes 
from $\alpha = 0.6$ to $\alpha = 0.5$. A more extreme mass dependence
of the activity level will enhance this effect. In general, we expect that $\alpha$ will be underestimated by $\ga 0.1$, if the
masses are estimated from \teff\ and activity is not taken into account. In addition, the peak of the mass function 
%(calculated in log-log form) 
could be underestimated by up to a factor of 2. For a more quantitative assessment of these effects a larger sample 
of \ha\ measurements for young stars and BDs is needed.

%{\bf [KGS: Probably need to acknowledge some recent efforts to determine the IMF based on careful screening of sources
%with low extinction/accretion (e.g. the ONC by Da Rio et al 2011).]} 

\section{Conclusions\label{conclude}}
In this paper, we have shown that there exists a correlation between the strength of \ha\ emission in active M-dwarfs, and the degree to which their temperatures are suppressed and radii inflated compared with inactive objects and theoretical evolutionary models.  By applying these relations, we are able to 
infer the amount by which an active objects' temperatures have been suppressed, and thereby
improve the accuracy of estimates for their masses and radii. 
We use the brown dwarf eclipsing binary system \2m\ as a benchmark for our model.
We expect these relations to be most useful for correcting
estimated masses and radii of low-mass stars and brown dwarfs over their
active lifetimes \citep[few Gyr;][]{west08} 
%\sout{and thereby also the inferred masses of objects with unknown} 
and when the ages or distances (and therefore the luminosities) are unknown. 
We have shown that failing to account for the effects of activity can cause significant errors in estimates of stellar and substellar masses derived from \teff, and smaller, but systematically biased errors in temperature and radius. 

%In summary, this analysis suggests that when using \teff\ to derive , the effects of magnetic activity
%may be significant.
%A correction based on the activity level is needed. 
%Although both young low-mass stars and BDs are the best candidates for applying our empirical correction, 
%this is
%difficult to achieve, because their activity signatures are contaminated (or in fact dominated) by accretion.
%To eliminate the effects of magnetic activity, \lbol\ can be used to derive stellar and substellar masses. The use of luminosities,
%however, is problematic as well for very young objects, because luminosities are strongly sensitive to age, accretion, disk
%excess, and extinction -- although careful evaluation of these parameters can mitigate these problems \citep[e.g.][]{dario12}.
%Taken together, these difficulties severely limit our ability to accurately derive the mass function for very
%young stars and brown dwarfs.

If these empirical corrections are corroborated by future observations, they will prove valuable not only for individual objects, but also for studies of stellar populations. In the case of individual objects, reliable stellar properties are invaluable for exoplanetary studies, where exquisite knowledge of the host star is required to infer planet properties.  For very young stellar populations, where activity levels are highest, we have shown how underestimated masses can substantially shift the inferred initial mass function.  Such a change would necessitate revisions to star formation models and population synthesis models because, e.g., the observed fraction of brown dwarfs and low-mass stars might be substantially altered.

While promising, the correlations we have derived contain significant scatter, and they are currently limited by the lack of a single sample of stars with both \ha\ and direct radius measurements.  A sample of \ha\ measurements for objects with directly measured radii and temperatures will allow us to better assess our relations and determine if the scatter is intrinsic or if it is caused by the intermediate steps necessary in constructing our relation (e.g., converting \lx\ to \lha).  Despite this limitation, we have chosen to pursue a correlation with \ha\ emission (rather than X-ray, or radio, for example) because of the relative ease of its measurement even in the substellar regime. In principle, a reliable activity tracer in the near-IR would prove even more useful by making measurements easier for cooler objects. No such tracer has yet been identified 
\citep[see, e.g.,][]{Schmidt2012}, although the \ion{He}{1} line at 10830\AA\ may be a possibility \citep[e.g.][]{dupree92}.

In order to make progress on understanding how \teff\ suppression and radius inflation relate to chromospheric activity, a larger sample of EBs with precise masses, radii, {\it and} \ha\ measurements are needed.  Toward this goal, we encourage other researchers to publish \ha\ measurements for their targets, as this is usually readily available from the spectra used to determine radial velocities.  
Indeed, a number of low-mass EBs with accurate masses and radii have been published in the last few years
\citep[e.g.][]{vaccaro07,irwin09,morales10,kraus11,helminiak11a,helminiak11b,irwin11},
potentially increasing by a factor of 2--3 the small sample that we have used from \citet{lopez07}. 
We are currently collecting new \ha\ measurements for these published EBs that lack resolved \ha\ measurements in order to improve the empirical relations that we have presented here.
%In addition, many of these may possess \lx\ measurements in the literature that could be used to infer \lha\ 
%as we have done above.

Finally, the relations we have determined already indicate quite clearly that the radius inflation and temperature
suppression mechanism operates in such a way that the temperature suppression and radius inflation almost 
exactly cancel in terms of their effect on the bolometric luminosity. 
Moreover, the relations between activity, \teff\ suppression, and radius inflation do not appear to manifest any
obvious discontinuity across the fully convective transition \citep[see also][]{stassun10};
the followup observations we have underway should help to refine this. 
These are important, fundamental clues to the
physical nature of these effects, and should help to constrain theoretical models that are being developed to 
explain these phenomena \citep[e.g.][]{chabrier07,macdonald2009}.

\acknowledgments
We thank Beate Stelzer for helpful discussions.
K.G.S.\ acknowledges NSF grants AST-0607773 and AST-1009810. 
Part of this work was funded by the Science Foundation Ireland through grant no.\ 
10/RFP/AST2780 to A.S. 
T.J.D.\ acknowledges support from Hubble Fellowship grant HST-HF-51271.01-A awarded by the Space
Telescope Science Institute, which is operated by AURA for NASA, under contract NAS 5-26555.

%\clearpage

%\clearpage

\end{document}